\pdfoutput=1
\documentclass[PAPER, atlasdraft=false, USenglish, cernpreprint, texmf, orcidlogo]{atlasdoc}
 
\usepackage{multirow}
\usepackage{lscape}
\usepackage{array}
\usepackage{cancel}
 
\usepackage{atlaspackage}[subfig]
\usepackage{atlasbiblatex}
 
\usepackage{atlasphysics}[snippets=true,jetetmiss=true]

\graphicspath{{logos/}{figures/}}
 
\usepackage{ANA-EXOT-2023-09-PAPER-defs}

\AtlasTitle{Search for dark mesons decaying to top and bottom quarks in proton--proton collisions at $\sqrt{s}=13~\TeV$ with the ATLAS detector}

\AtlasAbstract{
A search for dark mesons originating from strongly-coupled, $SU(2)$ dark flavor symmetry conserving models and decaying gaugephobically to pure Standard Model final states containing top and bottom quarks is presented. The search targets fully hadronic final states and final states with exactly one electron or muon and multiple jets. The analyzed data sample corresponds to an integrated luminosity of \lumi of proton--proton collisions collected at $\sqrt{s}=13~\TeV$ with the ATLAS detector at the Large Hadron Collider. No significant excess over the Standard Model background expectation is observed and the results are used to set the first direct constraints on this type of model. The two-dimensional signal space of dark pion masses $m_{\piD}$ and dark rho-meson masses $m_{\rho_D}$ is scanned. For $m_{\piD}/m_{\rho_D}=0.45$, dark pions with masses $\dpmass<940~\GeV$ are excluded at the $95\%$ CL, while for $m_{\piD}/m_{\rho_D}=0.25$ masses $\dpmass<740~\GeV$ are excluded.
}

\AtlasRefCode{EXOT-2023-09}
 
\PreprintIdNumber{CERN-EP-2024-150}

\AtlasJournal{JHEP}
\AtlasJournalRef{JHEP 09 (2024) 005}
\AtlasDOI{10.1007/JHEP09(2024)005}

\hypersetup{pdftitle={ATLAS document},pdfauthor={The ATLAS Collaboration}}
 
\begin{document}
 
\maketitle
 
\tableofcontents

\section{Introduction}
\label{sec:intro}

 
The Standard Model (SM) can be extended by a new strongly-coupled, confining gauge theory with fermion representation which transforms under the electroweak group.
The appeal of such an extension is that dark matter can arise in the form of composite mesons or baryons of the new strongly-coupled theory. In addition, these models often exhibit an automatic accidental symmetry protecting against dark matter decay.
Consequently, candidates for strongly-coupled dark matter include dark mesons, dark quarkonia-like states, glueballs and dark baryons~\cite{Kribs:2016cew,Butterworth:2021jto}.
The search presented here targets one set of models incorporating this concept: Stealth Dark Matter~\cite{Appelquist:2015yfa}.
Here, the new strongly-coupled dark sector consists of vector-like fermions that can transform under the new dark group but also interact with both the electroweak sector of the SM and the Higgs boson.
The result is the emergence of a quantum chromodynamics (QCD)-like dark sector as the direct analog to the QCD meson and baryon sector.
This leads to several intriguing phenomenological consequences:
as long as the vector-like mass is dominant over the chiral mass, the new dark sector is only weakly constrained by precision electroweak or Higgs coupling measurements, while the Higgs interactions break the dark sector global symmetry and thus allow dark mesons to decay into pure SM states~\cite{Kribs:2018ilo}.
This search focuses on the low-energy effective theories developed in Ref.~\cite{Kribs:2018oad}, which incorporate the leading interactions between dark mesons of a strongly-coupled $SU(2)$ dark flavor symmetry preserving dark sector and the SM.
These models contain a stable dark scalar baryon which could account for the stable dark matter observed in cosmological measurements~\cite{Appelquist:2015yfa}.
 
The simplified model targeted in this search contains only the two phenomenologically relevant sets of dark mesons: a lighter pseudoscalar triplet of dark pions, \piD, and an additional triplet of dark rho vector mesons, \rhoD, which are both expected at a scale around or slightly above the electroweak scale.
Following standard theoretical assumptions, the triplets are completely mass-degenerate and the dark sector can be fully described by three parameters: the mass of the dark pions $m_{\pi_D}$, the mass of the dark rho mesons $m_{\rho_D}$, and the number of dark colors $N_D$.
Since the phenomenological consequences remain unchanged for values of $N_D$ that are not excessively large, $N_D$ is fixed to $N_D=4$ throughout this search following the typical choice made for Stealth Dark Matter~\cite{Appelquist:2015yfa}.

\begin{figure}[tbp]
\centering
\subfloat[]{\includegraphics[page=6,width=0.31\textwidth]{fig_01a.pdf}\label{fig:FeynmanW}} \hspace{2mm}
\subfloat[]{\includegraphics[page=2,width=0.31\textwidth]{fig_01b.pdf}\label{fig:FeynmanDY}} \hspace{2mm}
\subfloat[]{\includegraphics[width=0.31\textwidth]{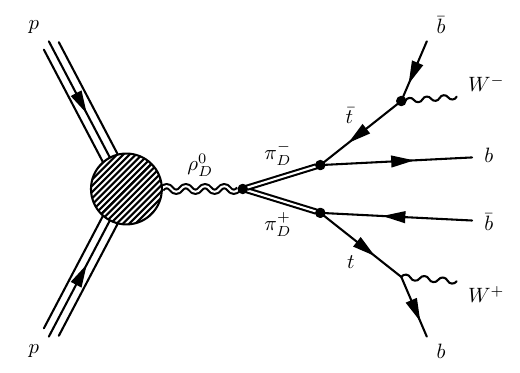}\label{fig:FeynmanDecay}} 
\caption{Examples of leading Feynman diagrams of dark pion pair production. The diagram in \protect\subref{fig:FeynmanW} shows resonant production via kinetic mixing with the \Wboson-field resulting in either a neutral or charged dark rho meson, a mixing with the $B$-field that can only result in a neutral dark rho meson is also possible, \protect\subref{fig:FeynmanDY} shows Drell--Yan-type pair production of dark pions, and \protect\subref{fig:FeynmanDecay} shows an illustrative diagram of the dark pion decay into a top and a bottom quark for dark pion production.}
\label{fig:feynmangraphs}
\end{figure}
 
Contrary to QCD, the vector-like nature of the dark sector allows to either gauge the full \sul weak interaction symmetry group or just the underlying $U(1)$ group, which leads to two distinct models of kinetic mixing of dark mesons with the SM, \sul and \sur.
The phenomenological consequences manifest themselves in the allowed decay channels and production cross-sections of dark pions, where the \sul models result in considerably larger cross sections than the \sur models.
Dark pions are always pair-produced either via Drell--Yan-type processes or resonantly via kinetic mixing of SM electroweak gauge bosons with the \rhoD that then subsequently decays into a pair of dark pions, as shown in
Figure~\ref{fig:feynmangraphs}. The kinetic mixing parameter $\epsilon$
depends on the number of dark colors as shown in Figure~\ref{fig:FeynmanW} (see also Ref.~\cite{Kribs:2018ilo}).
Throughout nearly all of the parameter space investigated in this search, the resonant production dominates the production of dark pions.
Once the choice of $N_D=4$ is made, the production cross-section depends trivially on the ratio of the dark pion and dark rho-meson masses, for which the symbol $\etaD=m_{\piD}/m_{\rhoD}$ is used, equivalent to the $\eta$ defined in Ref.~\cite{Kribs:2018ilo}. For gaugephobic \sul models, a given dark pion mass and \etaD-parameter fully specify the model, including the dark pion decay branching fractions.

This search considers only models with $\etaD<0.5$ where the decay $\rho_D^{\pm,0}\rightarrow\pi_D^{\pm}\pi_D^{0,\mp}$ has a branching fraction of nearly $1.0$, while for models with $\etaD>0.5$ this decay is kinematically forbidden and the dark rho meson can only decay to pairs of leptons or quarks.
Previous searches for resonances in the dilepton spectrum both in ATLAS~\cite{EXOT-2016-05} and CMS~\cite{CMS-EXO-16-047} have placed strong constraints on such models~\cite{Kribs:2018ilo}.
The bounds for models with $\etaD<0.5$ are considerably weaker~\cite{Butterworth:2021jto}.
This is the first search in any collider experiment optimized for this specific type of model. 
 
\begin{figure}[tbp]
\centering
\subfloat[]{\includegraphics[width=0.495\textwidth]{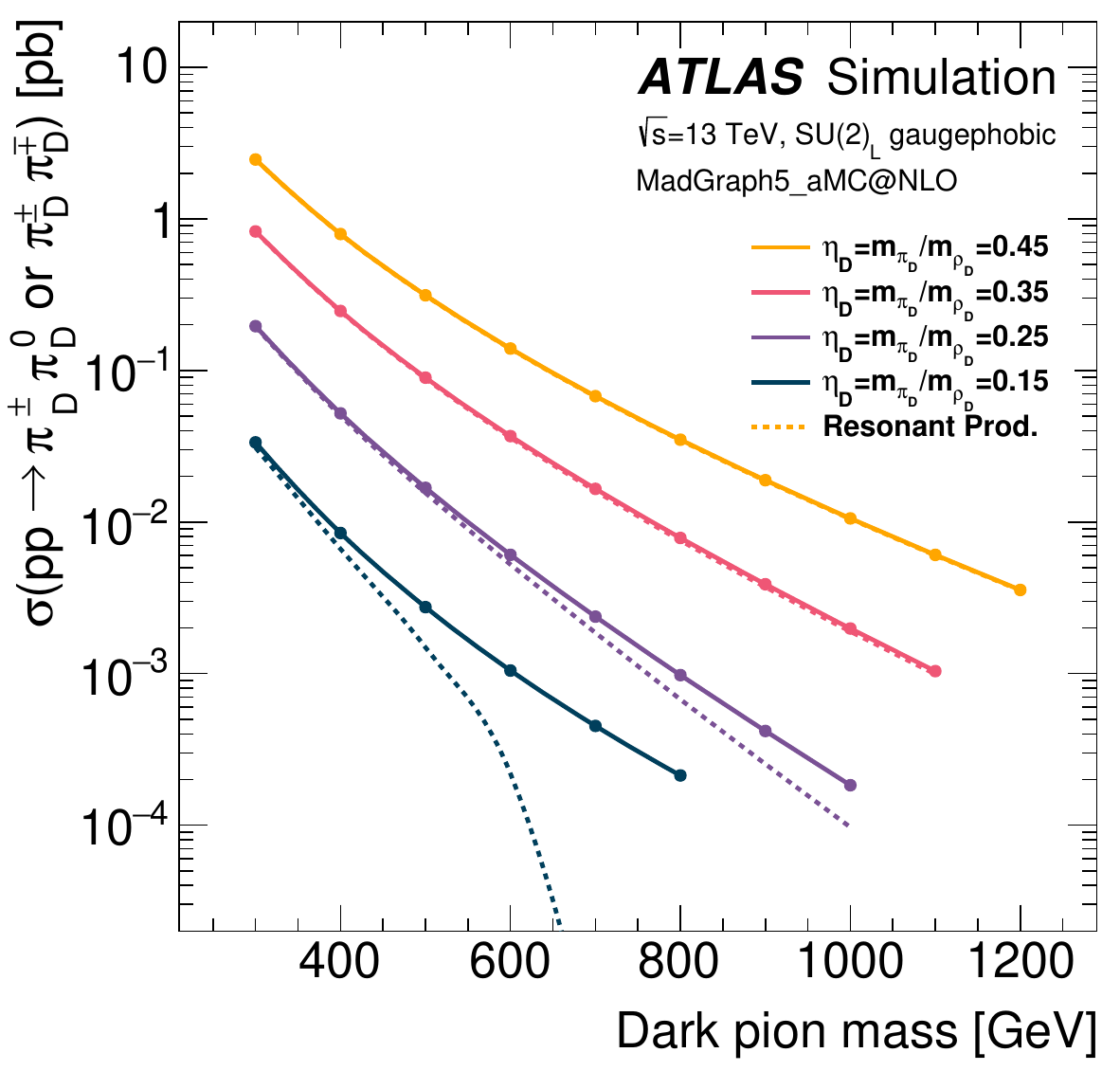}\label{fig:crosssections_su2l}}
\subfloat[]{\includegraphics[width=0.495\textwidth]{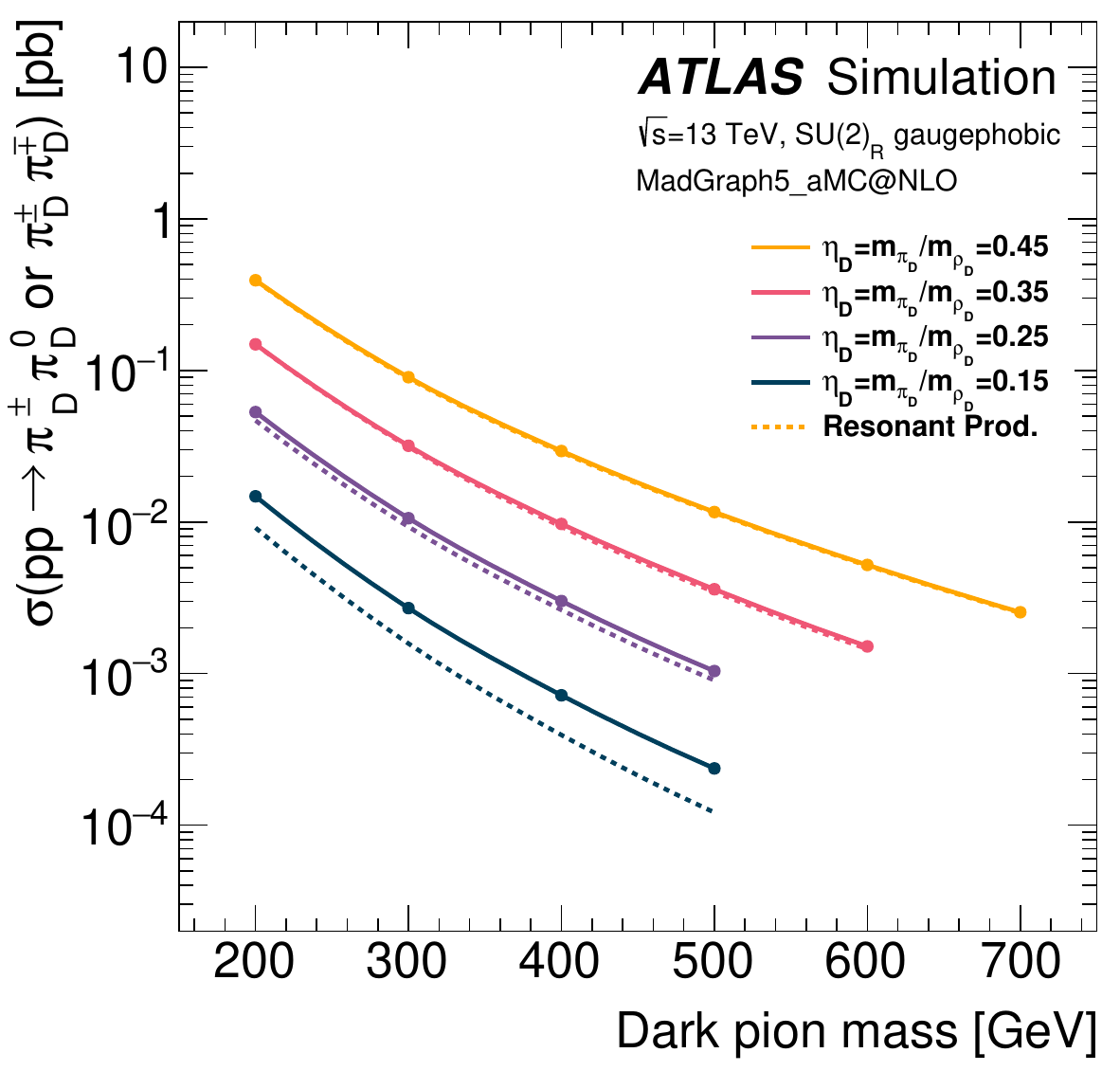}\label{fig:crosssections_su2r}}
\caption{Pair-production cross-sections for dark pions as a function of dark pion mass for four different values of $\etaD$ in \protect\subref{fig:crosssections_su2l} an \sul model and \protect\subref{fig:crosssections_su2r} an \sur model. The dashed colored lines indicate the contribution of the resonant production mode to the total dark pion production cross-section.}
\label{fig:crossections}
\end{figure}

\begin{figure}[tbp]
\centering
\subfloat[]{\includegraphics[page=3,width=0.495\textwidth]{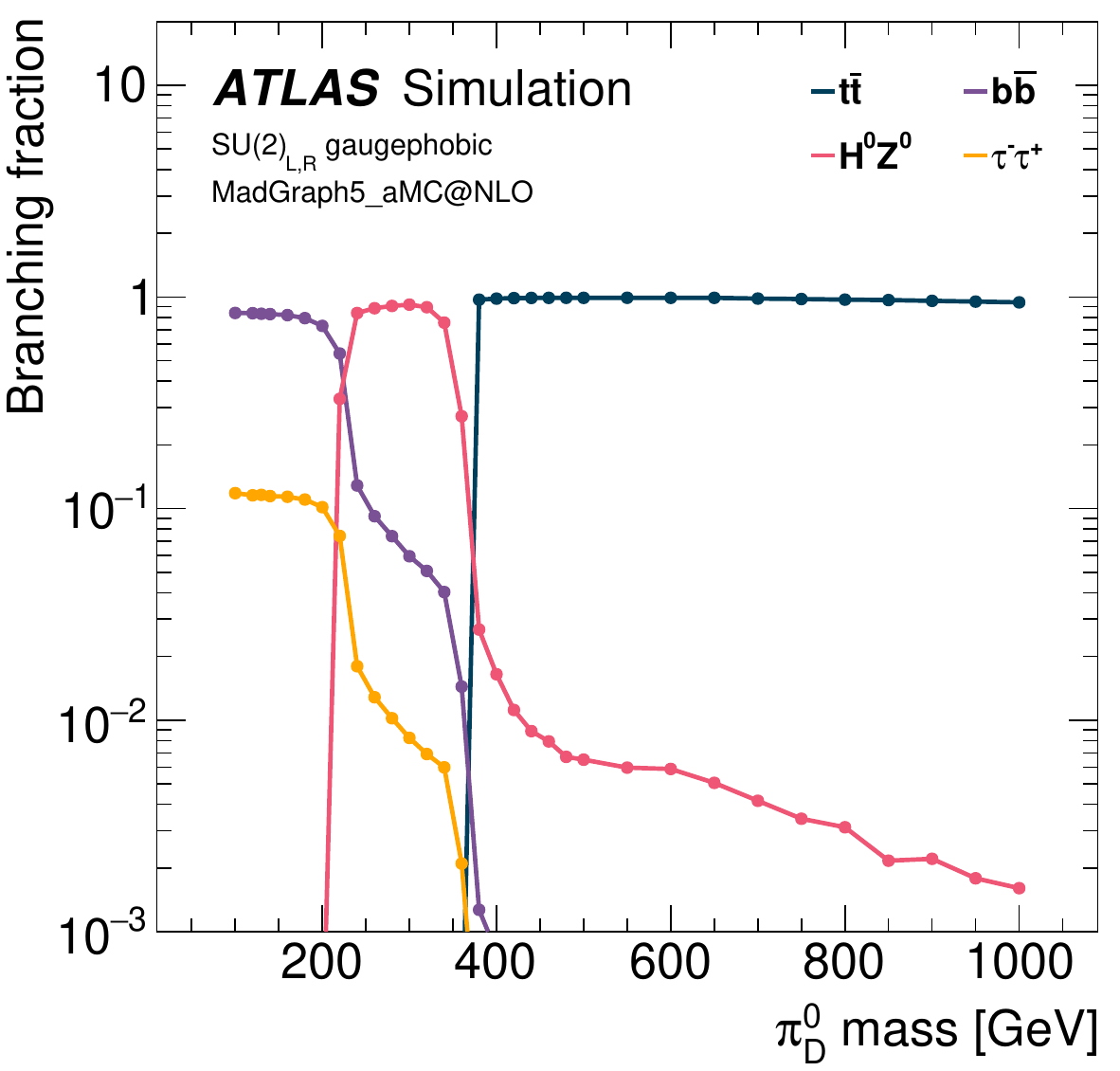}\label{fig:pi0branchingfraction}}
\subfloat[]{\includegraphics[page=4,width=0.495\textwidth]{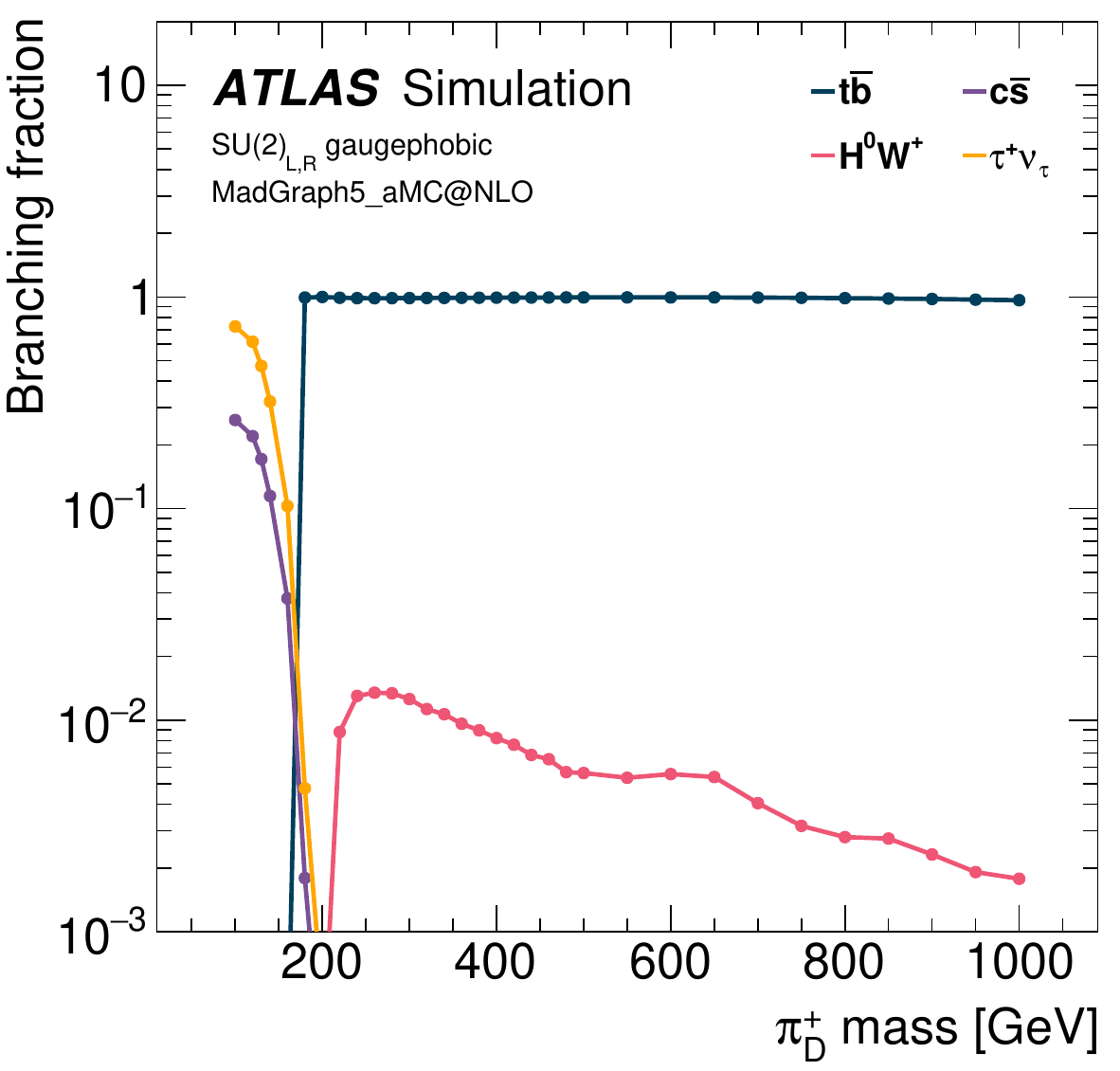}\label{fig:pipbranchingfraction}}
\caption{Branching fractions of the available decays of dark pions from gaugephobic \sul and \sur  models are shown for \protect\subref{fig:pi0branchingfraction} neutral dark pions and \protect\subref{fig:pipbranchingfraction} positively charged dark pions. Channels with small branching fractions are suppressed for clarity.}
\label{fig:branchingfractions}
\end{figure}

Figure~\ref{fig:crossections} shows the pair-production cross-sections for dark pions in \sul and \sur models.
The contribution of resonant production to the total production cross-section is indicated by the dashed lines.
A variety of different decay channels are open to dark pions in the available parameter space.
The most relevant channels and their branching fractions are shown in Figure~\ref{fig:branchingfractions}.
For gaugephobic models the decay to top and bottom quarks dominates at high masses, while decays to bottom and charm quarks, $\tau$-leptons and gauge bosons are relevant at lower dark pion masses.
 
This search is the result of the analysis of \lumi of proton–proton (\pp) collisions collected by the ATLAS detector during Run 2 of the Large Hadron Collider (LHC).  Since the dark pions are pair-produced in the model considered, the experimental signatures are either three top quarks and one bottom quark ($tttb$) or two top quarks and two bottom quarks ($ttbb$)\footnote{The label "$tttb$" is used to indicate both $t\bar{t}t\bar{b}$ as well as its charge conjugate, $t\bar{t}\bar{t}b$; whereas "$ttbb$" refers to $t\bar{t}b\bar{b}$.}. About one third of dark pions in the \sul models are neutral, resulting in the $tttb$ event signature being twice as likely as the $ttbb$ signature. These processes can give rise to several different final states depending on the hadronic
or semileptonic decay mode of each of the top quarks.
The search is performed in the \emph{\allhad} channel, targeting fully hadronic top quark decays where the signal results in eight to ten jets of which at least four originate from bottom quarks, and in the \emph{\onelep} channel, corresponding to final states with exactly one electron or muon in addition to up to four jets from b-quarks.
 
The results are interpreted as limits on the production cross-section of dark pion pairs as a function of $m_{\piD}$ and \etaD.


\section{ATLAS detector}
\label{sec:detector}

\newcommand{\AtlasCoordFootnote}{
ATLAS uses a right-handed coordinate system with its origin at the nominal interaction point (IP)
in the center of the detector and the \(z\)-axis along the beam pipe.
The \(x\)-axis points from the IP to the center of the LHC ring,
and the \(y\)-axis points upwards.
Polar coordinates \((r,\phi)\) are used in the transverse plane,
\(\phi\) being the azimuthal angle around the \(z\)-axis.
The pseudorapidity is defined in terms of the polar angle \(\theta\) as \(\eta = -\ln \tan(\theta/2)\) and is equal to the rapidity
$ y = \frac{1}{2} \ln \left( \frac{E + p_z c}{E - p_z c} \right) $ in the relativistic limit.
Angular distance is measured in units of \(\Delta R \equiv \sqrt{(\Delta y)^{2} + (\Delta\phi)^{2}}\).}

The ATLAS detector~\cite{PERF-2007-01} at the LHC covers nearly the entire solid angle around the collision point.\footnote{\AtlasCoordFootnote}
It consists of an inner tracking detector surrounded by a thin superconducting solenoid, electromagnetic and hadronic calorimeters,
and a muon spectrometer incorporating three large superconducting air-core toroidal magnets.
 
The inner-detector system (ID) is immersed in a \qty{2}{\tesla} axial magnetic field
and provides charged-particle tracking in the range \(|\eta| < 2.5\).
The high-granularity silicon pixel detector covers the vertex region and typically provides four measurements per track,
the first hit generally being in the insertable B-layer (IBL) installed before Run~2~\cite{ATLAS-TDR-19,PIX-2018-001}.
It is followed by the SemiConductor Tracker (SCT), which usually provides eight measurements per track.
These silicon detectors are complemented by the transition radiation tracker (TRT),
which enables radially extended track reconstruction up to \(|\eta| = 2.0\).
The TRT also provides electron identification information
based on the fraction of hits (typically 30 in total) above a higher energy-deposit threshold corresponding to transition radiation.
 
The calorimeter system covers the pseudorapidity range \(|\eta| < 4.9\).
Within the region \(|\eta|< 3.2\), electromagnetic calorimetry is provided by barrel and
endcap high-granularity lead/liquid-argon (LAr) calorimeters,
with an additional thin LAr presampler covering \(|\eta| < 1.8\)
to correct for energy loss in material upstream of the calorimeters.
Hadronic calorimetry is provided by the steel/scintillator-tile calorimeter,
segmented into three barrel structures within \(|\eta| < 1.7\), and two copper/LAr hadronic endcap calorimeters.
The solid angle coverage is completed with forward copper/LAr and tungsten/LAr calorimeter modules
optimized for electromagnetic and hadronic energy measurements respectively.
 
The muon spectrometer (MS) comprises separate trigger and
high-precision tracking chambers measuring the deflection of muons in a magnetic field generated by the superconducting air-core toroidal magnets.
The field integral of the toroids ranges between \num{2.0} and \qty{6.0}{\tesla\metre}
across most of the detector.
Three layers of precision chambers, each consisting of layers of monitored drift tubes, cover the region \(|\eta| < 2.7\),
complemented by cathode-strip chambers in the forward region, where the background is highest.
The muon trigger system covers the range \(|\eta| < 2.4\) with resistive-plate chambers in the barrel, and thin-gap chambers in the endcap regions.
 
The luminosity is measured mainly by the LUCID--2~\cite{LUCID2} detector that records Cherenkov light produced in the quartz windows of photomultipliers located close to the beampipe.
 
Events are selected by the first-level trigger system implemented in custom hardware,
followed by selections made by algorithms implemented in software in the high-level trigger~\cite{TRIG-2016-01}.
The first-level trigger accepts events from the \qty{40}{\MHz} bunch crossings at a rate below \qty{100}{\kHz},
which the high-level trigger further reduces in order to record complete events to disk at about \qty{1}{\kHz}.
 
A software suite~\cite{SOFT-2022-02} is used in data simulation, in the reconstruction
and analysis of real and simulated data, in detector operations, and in the trigger and data acquisition
systems of the experiment.


\section{Data and simulated event samples}
\label{sec:samples}

 
This analysis is performed using data from \pp collisions with $\rts=13\,\TeV$ recorded by the ATLAS detector in 2015--2018. Only events for which all relevant subsystems were operational are considered. The data correspond to an integrated luminosity of $140.1\pm1.2\,\ifb$~\cite{DAPR-2021-01}.
 
Monte Carlo (MC) simulated event samples are used for the estimate of background from SM processes and to model the targeted signal models. The details of the event generation are provided in Sections~\ref{sec:samples:sig}~and~\ref{sec:samples:bkg} for signal and background samples, respectively.
The generation of all simulated event samples includes the effect of multiple \pp interactions per bunch crossing, as well as changes in detector response due to interactions in bunch crossings before or after the one containing the hard interaction, modeled by overlaying simulated inelastic events on the physics event. These two effects are referred to as pileup. The simulated event samples are processed with the \GEANT-based ATLAS detector simulation~\cite{Agostinelli:2002hh,SOFT-2010-01}.
All samples are weighted to match the pileup distribution observed in data and are processed with the same reconstruction algorithms as data.
 
\subsection{Signal samples}
\label{sec:samples:sig}
Signal samples are generated in a grid over a two-dimensional space, varying the dark pion mass $\dpmass$ between $300-1200~\GeV$ and the $\etaD$-parameter between $0.15-0.45$.
The matrix element calculation for the pair production of dark pions is performed at next-to-leading order~(NLO) in QCD based on the model described in Ref.~\cite{Kribs:2018ilo} using \mgamc~v2.4.3~\cite{Alwall:2014hca} interfaced with \pythia.212~\cite{Sjostrand:2014zea} for the modeling of parton showering, hadronization and underlying event using the A14 set of tuned parameters ("tune")~\cite{ATL-PHYS-PUB-2014-021} and the \nnpdftwo~\cite{Ball:2012cx} set of parton distribution functions (PDF).
 
The decays of bottom and charm hadrons are simulated using the \evtgen\ v1.6.0 program~\cite{Lange:2001uf}.
An additional set of signal samples, with parameter values near the expected exclusion contour of the \allhad channel, is generated at NLO in QCD using \mgamc~v2.9.9~\cite{Alwall:2014hca} interfaced with \pythia.306~\cite{AComGuiToThe} using the A14 tune and \nnpdftwo PDF set. Kinematic distributions match in both setups. All signal cross-sections are extracted from \mgamc~v2.9.9. The number of dark colors $N_D$ is set to $4$ for all signal points. The dark pion decays are simulated using the narrow width approximation and contain all possible decay channels from Figure~\ref{fig:branchingfractions}.
 
\subsection{Background samples}
\label{sec:samples:bkg}
The dominant SM background process in the \allhad channel is multijet production. This background is estimated with data-driven methods while MC simulation is used to estimate the remaining SM processes. The background in the \onelep channel is estimated from MC simulations and is dominated by top quark pair-production (\ttbar), often in association with heavy-flavor quarks (\ttHF). Other important backgrounds are the production of a vector boson in association with jets (\Vjets) and single top-quark production (single top) which is dominated by the associated production of a top quark with a \Wboson boson but also contains single top-quark production in the $s-$ and $t-$channels. Smaller background contributions stem from \ttbar produced in association with additional bosons or quarks (\tttt, \ttV, \ttH, and other \ttX) and multiboson production.
The configurations used to produce the background samples are described below and are summarized in Table~\ref{tab:bkgsamples}. For all background samples, except those generated with \sherpa, the \evtgen\ v.1.6.0 or v1.7.0 program is used to simulate the decays of bottom and charm hadrons.
 
\begin{table}[tbp]
\caption{Overview of the configuration of all nominal background samples used in the analysis; details and definitions are provided in the text.}
\label{tab:bkgsamples}
\centering
\resizebox{\textwidth}{!}{\begin{tabular}{l|lllll}
\toprule
Process & Generator & PDF & Showering & Tune & Cross section \\
\midrule
\ttbar & \powhegbox~v2 & \nnpdfnlo & \pythia & \textsc{A14} & \textsc{NNLO+NNLL} \\ 
\ttbb & \powhegboxres & \nnpdfnlo & \pythia & \textsc{A14} & \textsc{NLO} \\ 
\Vjets & \sherpa~v2.2.11 & \nnpdfnnlo & \sherpa & Def. & \textsc{NLO} \\ 
Single top & \powhegbox~v2 & \nnpdfnlo & \pythia & \textsc{A14} & \textsc{NLO+NNLL} \\ 
\tttt & \mgamc~v2.4.3 & \nnpdfonenlo & \pythia & \textsc{A14} & \textsc{NLO} \\ 
\ttV & \mgamc~v2.3.3 & \nnpdfnlo & \pythia & \textsc{A14} & \textsc{NLO} \\ 
\ttH & \powhegbox~v2 & \nnpdfnlo & \pythia & \textsc{A14} & \textsc{NLO} \\ 
Other \ttX & \mgamc & \nnpdftwo & \pythia & \textsc{A14} & \textsc{NLO} \\
Multiboson & \sherpa~v2.2.1/v2.2.2 & \nnpdfnnlo & \sherpa & Def. & \textsc{NLO} \\
\bottomrule
\end{tabular}}
\end{table}
 
\subsubsection{\ttbar background}
\label{sec:ttbarBackground}
The production of \ttbar events is modeled using the \powhegbox~v2~\cite{Frixione:2007nw,Nason:2004rx,Frixione:2007vw,Alioli:2010xd} generator that provides matrix elements at NLO in QCD with the \nnpdfnlo~\cite{Ball:2014uwa,Carrazza:2015aoa} set PDFs and the \hdamp\ parameter, which controls the matching in \powheg and effectively regulates the high-\pt radiation against which the \ttbar system recoils, set to 1.5~\mtop~\cite{ATL-PHYS-PUB-2016-020}.
The functional form of the renormalization and factorization scales are set to the default scale $\sqrt{m_{\textrm{top}}^2 + p_{\textrm T}^2}$.
The events are interfaced with \pythia.230 for the parton shower and hadronization, using the A14 set of tuned parameters and the \nnpdftwo PDF set.
The \ttbar sample is normalized to the cross section prediction at next-to-next-to-leading order (NNLO) in QCD including the resummation of next-to-next-to-leading-logarithmic (NNLL) soft-gluon terms calculated using \textsc{Top++2.0}~\cite{Beneke:2011mq,Cacciari:2011hy,Baernreuther:2012ws,Czakon:2012zr,Czakon:2012pz,Czakon:2013goa,Czakon:2011xx}.
For \pp collisions at a center-of-mass energy of $\sqrt{s}=13~\TeV$, this cross section corresponds to $\sigma(\ttbar)_{\textrm{NNLO+NNLL}} = \mathrm{832\pm51~pb}$ using a top-quark mass of $\mtop = 172.5~\GeV$.
 
The inclusive \ttbar sample described above is complemented by a dedicated sample in which a pair of top quarks is produced in association with two $b$-quarks. Events are simulated with the \powhegboxres~\cite{Jezo:2018yaf} generator and \openloops~\cite{Buccioni:2019sur,Cascioli:2011va,Denner:2016kdg}, using a pre-release of the implementation of this process in \powhegboxres provided by the authors~\cite{ttbbPowheg}, with the \nnpdfnlo PDF set. It is interfaced with \pythia.240, using the A14 set of tuned parameters and the \nnpdftwo PDF set. The four-flavor scheme is used with the $b$-quark mass set to $4.95~\GeV$. The factorization scale and \hdamp parameter are both set to $0.5\times\Sigma_{i=t,\bar{t},b,\bar{b},j}m_{\mathrm{T},i}$, and the renormalization scale is set to $\sqrt[4]{m_{\text{T}}(t)\cdot m_{\text{T}}(\bar{t})\cdot m_{\text{T}}(b)\cdot m_{\text{T}}(\bar{b})}$. This \ttbb sample is used in the \onelep channel where the dominant background comes from \ttbar production, whereas the multijet-dominated \allhad channel relies on the five-flavor scheme inclusive sample alone.
 
Previous studies have seen improved agreement between data and prediction in \ttbar events, particularly for the top-quark \pt distribution, when comparing with NNLO calculations~\cite{TOPQ-2015-06}. Top-quark pair differential calculations at NNLO QCD accuracy including electroweak (EW) corrections have become available~\cite{Czakon:2017wor}. Hence, a small improvement to the modeling is incorporated by correcting the \ttbar and the \ttbb samples to match their top/antitop-quark \pt and the top-quark mass distribution to the accuracy predicted at NNLO in QCD and NLO in EW.
 
Events in the \ttbar and \ttbb samples are classified according to the flavor of the particle jets not originating from the top quark. The particle jets are
reconstructed from the simulated stable particles using the \antikt algorithm~\cite{Cacciari:2008gp, Fastjet} with a radius parameter $R=0.4$, and are required to have $\pt>15~\GeV$ and $|\eta|<2.5$.
Events are labeled as \ttb if at least one particle jet is matched within $\Delta R<0.4$ to $b$-hadrons with $\pt>5~\GeV$ that do not arise from the decay of top quarks. 
In the remaining events, if at least one particle jet is matched within $\Delta R<0.4$ to additional $c$-hadrons with $\pt>5~\GeV$, the events are classified as \ttc. All other events are labeled as \ttlight. The \ttbb sample is used for the \ttb category meaning that all \ttlight and \ttc events are rejected from this sample. Likewise, only the \ttlight and \ttc events are retained from the inclusive \ttbar sample.
 
\subsubsection{Other backgrounds}
The production of \Vjets is simulated with the \sherpa~v2.2.11~\cite{Bothmann:2019yzt} generator using NLO matrix elements for up to two partons, and leading-order (LO) matrix elements for up to five partons calculated with the Comix~\cite{Gleisberg:2008fv} and \openloops libraries.
They are matched with the \sherpa parton shower~\cite{Schumann:2007mg} using the MEPS@NLO prescription~\cite{Hoeche:2011fd,Hoeche:2012yf,Catani:2001cc,Hoeche:2009rj} and the set of tuned parameters developed by the \sherpa authors.
The \textsc{Hessian} \nnpdfnnlo PDF set is used and the samples are normalized to a prediction that is NNLO in QCD~\cite{Anastasiou:2003ds}.
 
The associated production of a top quark with a $W$ bosons ($tW$) is modeled using the \powhegbox~v2~\cite{Re:2010bp,Nason:2004rx,Frixione:2007vw,Alioli:2010xd} generator at NLO in QCD using the five-flavor scheme and the \nnpdfnlo PDF set.
The diagram removal scheme~\cite{Frixione:2008yi} is used to remove interference and overlap with \ttbar production.
The related uncertainty is estimated by comparing with an alternative sample generated using the diagram subtraction scheme~\cite{Frixione:2008yi,ATL-PHYS-PUB-2016-020}.
Single top-quark $t$-channel production is modeled using the \powhegbox~v2~\cite{Frederix:2012dh,Nason:2004rx,Frixione:2007vw,Alioli:2010xd} generator at NLO in QCD using the four-flavor scheme and the corresponding \nnpdfnlo PDF set.
Single top-quark $s$-channel production is modeled using the \powhegbox~v2~\cite{Alioli:2009je,Nason:2004rx,Frixione:2007vw,Alioli:2010xd} generator at NLO in QCD in the five-flavor scheme with the \nnpdfnlo PDF set.
All single top-quark events are processed through \pythia.230 using the A14 tune and the \nnpdftwo PDF set.
 
The production of \tttt events is modeled using the \mgamc~v2.4.3 generator which provides matrix elements at NLO in QCD with the \nnpdfonenlo~\cite{Ball:2014uwa} PDF set.
The functional form of the renormalization and factorization scales is set to 0.25$\times \sum_i \sqrt{m^2_i+p^2_{\text{T},i}}$, where the sum runs over all the particles generated from the matrix element calculation, following Ref.~\cite{Frederix:2017wme}.
Top quarks are decayed at LO using \madspin~\cite{Frixione:2007zp,Artoisenet:2012st} to preserve all spin correlations.
The events are interfaced with \pythia.230 for the parton shower and hadronization, using the A14 set of tuned parameters and the \nnpdftwo PDF set.
 
The production of \ttV events is modeled using the \mgamc~v2.3.3 generator at NLO in QCD with the \nnpdfnlo PDF set.
The events are interfaced to \pythia.210~ using the A14 tune and the \nnpdftwo PDF set.
 
The production of \ttH events is modeled using the \powhegbox~v2~\cite{Frixione:2007nw,Nason:2004rx,Frixione:2007vw,Alioli:2010xd,Hartanto:2015uka} generator at NLO in QCD with the \nnpdfnlo PDF set.
The events are interfaced to \pythia.230 using the A14 tune and the \nnpdftwo PDF set.
 
Further rare top-quark-pair backgrounds \ttt, \ttZZ, \ttWW, \ttWZ, \ttWH and \ttHH are all produced using the NLO in QCD \mgamc generator interfaced with \pythia using the A14 set of tuned parameters and scaled to NLO cross sections~\cite{deFlorian:2016spz}.
 
Samples of diboson final states ($VV$) are simulated with the \sherpa~v2.2.1 or v2.2.2~\cite{Bothmann:2019yzt} generator depending on the process, including off-shell effects and Higgs-boson contributions, where appropriate.
Semileptonic final states, where one boson decays leptonically and the other hadronically, are generated using matrix elements at NLO accuracy in QCD for up to one additional parton and at LO accuracy for up to three additional parton emissions.
Samples for the loop-induced processes $gg \to VV$ are generated using LO-accurate matrix elements for up to one additional parton emission.
The matrix element calculations are matched and merged with the \sherpa parton shower based on Catani--Seymour dipole factorization~\cite{Gleisberg:2008fv,Schumann:2007mg} using the MEPS@NLO prescription.
The virtual QCD corrections are provided by the \openloops library.
The \nnpdfnnlo PDF set is used, along with the dedicated set of tuned parton-shower parameters developed by the \sherpa authors.


\section{Object and event selections}
\label{sec:reco}


For each event, interaction vertices are reconstructed from ID charged particle tracks, where the tracks are required to have transverse momenta (\pt) greater than $500~\MeV$~\cite{ATL-PHYS-PUB-2015-026}. Candidates for the primary vertex are required to have at least two associated tracks. If multiple vertices are reconstructed, the vertex with the largest sum of the squares of the transverse momenta of associated tracks is taken as the primary vertex. Events that fail the primary vertex reconstruction are rejected.
 
Electrons are reconstructed from energy deposits in the electromagnetic calorimeter that are matched to charged-particle tracks in the ID~\cite{EGAM-2018-01}. They are identified using a likelihood-based (LH) identification which employs calorimeter and tracking information to discriminate between electrons and jets and that combines this likelihood and the likelihood of it originating from background processes into a single discriminant. Only electron candidates with $\pt>10~\GeV$ within $|\eta|< 1.37$ or $1.52 < |\eta| < 2.47$ are considered. 
Electrons are required to be well isolated using criteria based on the properties of the topological clusters in the calorimeter and of ID tracks around the reconstructed electron. Further requirements of $|z_0\sin\theta|<0.5~$mm and $|d_0|/\sigma(d_0)<5$ are placed on the longitudinal and transverse impact parameters to select electrons originating from the primary vertex.
Electrons are further categorized as “baseline” or “signal”. For the \allhad channel, baseline electrons are identified by the \emph{LooseAndBLayer} likelihood-based identification working point and are required to fulfill the \emph{Loose} isolation criteria~\cite{PERF-2017-01,EGAM-2018-01}. Events containing a baseline electron candidate satisfying these baseline criteria are rejected.
For the \onelep channel, baseline electrons are identified with the \emph{Medium} working point~\cite{EGAM-2018-01} and are not subject to any isolation requirement. Signal electrons are identified by the \emph{Tight} working point and are subject to the \emph{Tight} track-based isolation criteria~\cite{EGAM-2018-01}. The \pt-requirement of the signal electrons is increased to $\pt>28~\GeV$. Signal electrons constitute a subset of the baseline electrons.
 
Muon candidates are reconstructed by combining tracks in the MS with tracks in the ID and are subject to cut-based identification criteria which are based on the numbers of hits in the different ID and MS subsystems, and on the significance of the charge-to-momentum ratio $q/p$~\cite{MUON-2018-03}. All muon candidates are required to be within the acceptance region of the ID at $|\eta| < 2.5$ and to have $\pt>10~\GeV$. Muons are required to satisfy isolation requirements based on the properties of ID tracks around the reconstructed muon~\cite{MUON-2018-03}. Similarly to electrons, requirements on the longitudinal and transverse impact parameters, $|z_0\sin\theta|<0.5~$mm and $|d_0|/\sigma(d_0)<3$, are also applied.
Baseline muons are identified in the \allhad channel by the \emph{Loose} quality working point and are required to fulfill the \emph{Loose} isolation criteria~\cite{MUON-2018-03}. Events containing a muon candidate satisfying these baseline criteria are rejected. In the \onelep channel, baseline muons are identified by the \emph{Medium} quality working point~\cite{MUON-2018-03} and are not subject to any isolation criteria.
For the selection of signal muons the \emph{Medium} quality working point is applied and the muon candidates are required to fulfill the \emph{Tight} isolation criteria based on the $p_T^{\text{varcone30}}$ variable defined in Ref.~\cite{MUON-2018-03} and have $\pT>28~\GeV$. Signal muons constitute a subset of the baseline muons. Events are selected for the \onelep channel if they contain exactly one signal and no additional baseline leptons (electrons or muons).
 
Jet candidates are reconstructed using a particle-flow reconstruction algorithm~\cite{PERF-2015-09} combining charged particle tracks from the ID and three-dimensional topological energy clusters~\cite{PERF-2014-07} in the calorimeter. 
Jets are reconstructed using the \antikt\ algorithm~\cite{Cacciari:2008gp} implemented in the FastJet package~\cite{Fastjet} with a fixed radius parameter $R=0.4$ using charged constituents associated with the primary vertex and neutral particle flow objects as inputs.
In order to minimize the contribution from pileup jets, a requirement on the jet-vertex tagger~\cite{PERF-2014-03} is made for jets with \pt below 60~\GeV.
 
Slight differences in the efficiency of the association of jets to vertices in data and simulation are addressed by applying scale factors to simulation. Jet energy scale corrections, derived from MC simulation, are used to calibrate the average energies of jet candidates to the scale of their constituent particles~\cite{JETM-2018-05}. Remaining differences between data and simulated events are evaluated and corrected for using in situ techniques, which exploit the transverse momentum balance between a jet and a reference object such as a photon, \Zboson\ boson, or multijet system in data. After these calibrations, all jets in the event with $\pT > 20~\GeV$ must satisfy a set of loose jet-quality requirements~\cite{DAPR-2012-01} designed to reject jets originating from sporadic bursts of detector noise, large coherent noise or isolated pathological cells in the calorimeter system, hardware issues, beam-induced background or cosmic-ray muons~\cite{ATLAS-CONF-2015-029}.
In the \onelep channel, the jets are required to satisfy $|\eta|<2.5$, while in the  \allhad channel, they are required to satisfy $|\eta|<2.8$ to match the $\eta$ range of the $H_T$ trigger. If these jet requirements are not met, the jet is discarded.
 
Jets are tagged as containing a $b$-hadron ($b$-tagged) by a deep neural network algorithm trained on a simulated hybrid sample composed of \ttbar and $Z'\rightarrow\ttbar$ events~\cite{FTAG-2019-07,ATL-PHYS-PUB-2017-013} at a working point corresponding to a $77\%$ \bjet efficiency, as measured on an inclusive \ttbar sample. This working point has a rejection factor of $5$ and $170$ on charm and light-flavored jets, respectively.
Flavor-tagging efficiency differences between data and simulation are corrected by a reweighting procedure detailed in Refs.~\cite{FTAG-2018-01,FTAG-2019-02,FTAG-2020-08}.
 
To resolve any reconstruction ambiguities between electrons, muons and jets, an overlap removal procedure is applied in a prioritized sequence, based on baseline leptons and jets, as follows. First, if a electron shares the same ID track with another electron, the electron with lower \pt is discarded, and then any electron sharing the same ID track with a muon is rejected.
Next, jets are rejected if they lie within $\Delta R = 0.2$ of an electron. Similarly, jets within $\Delta R = 0.2$ of a muon are rejected if the jet has fewer than three associated tracks or if the muon is matched to the jet through ghost association~\cite{Cacciari:2008gp}. Finally, electrons that are close to a remaining jet are discarded if their distance from the jet is $\Delta R < 0.4$, while for muons the distance is $\Delta R < \text{min}(0.4, 0.04 + 10~\GeV/\pt)$.
 
Large-$R$ jets are reclustered from the calibrated $R=0.4$ jets described above using the \antikt algorithm with a fixed radius parameter of $R=1.2$~\cite{Nachman_2015}. The large-$R$ jets aim to fully contain the dark pion decay products and the $R$ parameter is optimized for the dark pion mass range of the targeted signal points.
For the \onelep channel the signal lepton is added to the $R=0.4$ jet collection before the reclustering, which then proceeds in the same way as in the \allhad channel. After reclustering, the large-$R$ jet containing the lepton is referred to as \jlep and the leading fully hadronic large-$R$ jet \jhad. Both large-$R$ jets originate from the same reclustering procedure, ensuring there is no overlap between the two.
 
Events are selected for the \allhad channel using triggers on \Ht, defined as the scalar sum of the transverse momenta of all the reconstructed jets in the event with $|\eta_{\text{jet}}|<2.8$~\cite{TRIG-2012-01}. The \Ht-trigger threshold was $850~\GeV$ in 2015 and the first half of 2016, and was increased to $1000~\GeV$ in the latter half of 2016 for the remainder of Run~2. Since the trigger decision is based on \Ht computed from trigger-level jet momenta (which lack a detailed calibration), the triggers show a slow onset behavior with respect to \Ht computed from jet momenta of fully calibrated jets. A requirement of $\Ht > 1150~\GeV$ ensures that the trigger is fully efficient to minimize systematic uncertainties resulting from imprecise modeling of the onset behavior in simulation. The \Ht variable computed from fully calibrated jets is used for the remainder of this search.
 
Events are selected for the \onelep channel using a combination of single-lepton triggers~\cite{TRIG-2018-01,TRIG-2018-05,TRIG-2019-04,TRIG-2019-03}. The single-lepton triggers require the presence of a muon or an electron with \pt higher than a certain threshold and, in some cases, impose identification and lepton-isolation requirements. The lowest \pt threshold was 24 (20)~\GeV for electrons (muons) during the 2015 data-taking period and 26~\GeV for both electrons and muons in the data-taking periods from 2016 to 2018.
The efficiencies of the single-lepton triggers range between $20\%$ and $50\%$ in the simulated signal samples. To account for small differences in the single-lepton trigger efficiency between data and simulation, all triggered simulated events receive an event weight to match data.
 
The analysis strategy relies on the reconstruction of each dark pion
using a large-$R$ jet.
In the \allhad channel, events are required to have six or more $R=0.4$ jets with $\pt>25~\GeV$, in addition to the $\Ht > 1150~\GeV$ requirement and the lepton veto. At least three jets within the ID acceptance ($|\eta| < 2.5$) must be $b$-tagged. At least two large-$R$ jets with jet mass $\makttwelve>190~\GeV$ are required in all events. Events for the \onelep channel are required to have at least five jets, out of which at least three have to be $b$-tagged, and to have $\Ht>300~\GeV$. Here the \Ht is defined similar to the \allhad channel, however only jets with $|\eta_{\text{jet}}|<2.5$ are considered. Events passing all selection requirements listed here are considered preselected for the analysis. The preselection requirements are summarized for both the analysis channels in Table~\ref{tab:preselection}.
 
\begin{table}[]
\centering
\caption{Summary of the preselection criteria for the \allhad and \onelep channels, in terms of the number of baseline and signal leptons, $R=0.4$ and $R=1.2$ jets, number of \bjets, and \Ht. The definitions of the physics objects for the two channels are given in the text. Signal leptons constitute a subset of the baseline leptons.
}
\label{tab:preselection}
\resizebox{0.5\textwidth}{!}{
\begin{tabular}{@{}l|cc@{}}
\toprule
Variable            & All-hadronic channel & One-lepton channel \\ \midrule
\nlep(baseline)     & $0$               & $1$              \\
\nlep(signal)       & -                 & $1$              \\
\njets($R=0.4$)     & $\geq 6$          & $\geq 5$         \\
\njets($R=1.2$)     & $\geq 2$          & -                \\
\nbjets             & $\geq 3$          & $\geq 3$         \\
\Ht                 & $\geq 1150~\GeV$ & $\geq 300~\GeV$ \\ \bottomrule
\end{tabular}
}
\end{table}


\section{Analysis strategy}
\label{sec:evtsel}

 
Preselected events are separated into signal, control, and validation regions based on the properties of the large-$R$ and small-$R$ jets in the event, as well as the signal lepton for the \onelep channel. Signal regions (SRs) are signal-enriched regions while the control regions (CRs) are used to estimate the SM background contributions. The validation regions (VRs) are used to validate the background estimation methods. The analysis strategies for the \allhad and \onelep channels are detailed below.

\subsection{All-hadronic channel}
\label{sec:evtsel_allhad}
In the \allhad channel, the leading two large-$R$ jets define the overall SR, where the leading large-$R$ jet satisfies $m_{\text{jet,R=1.2}}>300~\GeV$ and the sub-leading large-$R$ jet satisfies $m_{\text{jet,R=1.2}}>250~\GeV$ as shown in Figure~\ref{fig:rcjet_masses} for distributions after preselection. To suppress multijet events containing, for example, gluon to $\bbbar$ splitting, a selection on \mptbb, defined as the ratio of the mass to the transverse momentum of the pair of $b$-tagged jets closest to the center of the large-$R$ jet, is applied to both large-$R$ jets. In signal events \mptbb is expected to take on larger values than in background events, thus a cut of $\mptbb > 0.25$ is required. Further, both large-$R$ jets must satisfy a $bb_i$ tag, where the $\Delta R$ between the leading ($i=1$) or sub-leading ($i=2$) large-$R$ jet and the second closest $b$-tagged jet, defined as $\drbj$, is less than $1.0$ and thus both $b$-tagged jets are well contained within the volume of the large-$R$ jet. This variable is designed to suppress \ttbar events where the second closest \bjet arises from the other top quark in the event and can thus have a large $\Delta R$ with the large-$R$ jet. In signal, on the other hand, the decay products of the dark pion always include two $b$-quarks no matter whether the $tttb$ or $ttbb$ final state is considered and the $\Delta R$ therefore tends to be small. The overall SR is then subdivided into nine separate bins in the leading versus sub-leading large-$R$ jet mass plane.
A large-$R$ jet is considered $\pitagi$ tagged if its mass falls into one of the nine mass bins. The SR requires both large-$R$ jets to satisfy both tagging selections (i.e. both jets must be $bb_i$ and $\pitagi$ tagged). The events where the two leading large-$R$ jets satisfy only one or two out of the four possible tags form the CRs used for the data-driven multijet extrapolation; events that satisfy three tags allow for a validation of the method and thus form the VRs. The SR selection criteria are summarized in Table~\ref{tab:SR_tags}.
 
\begin{figure}[tb]
\centering
\subfloat[]{ \includegraphics[width=0.495\textwidth]{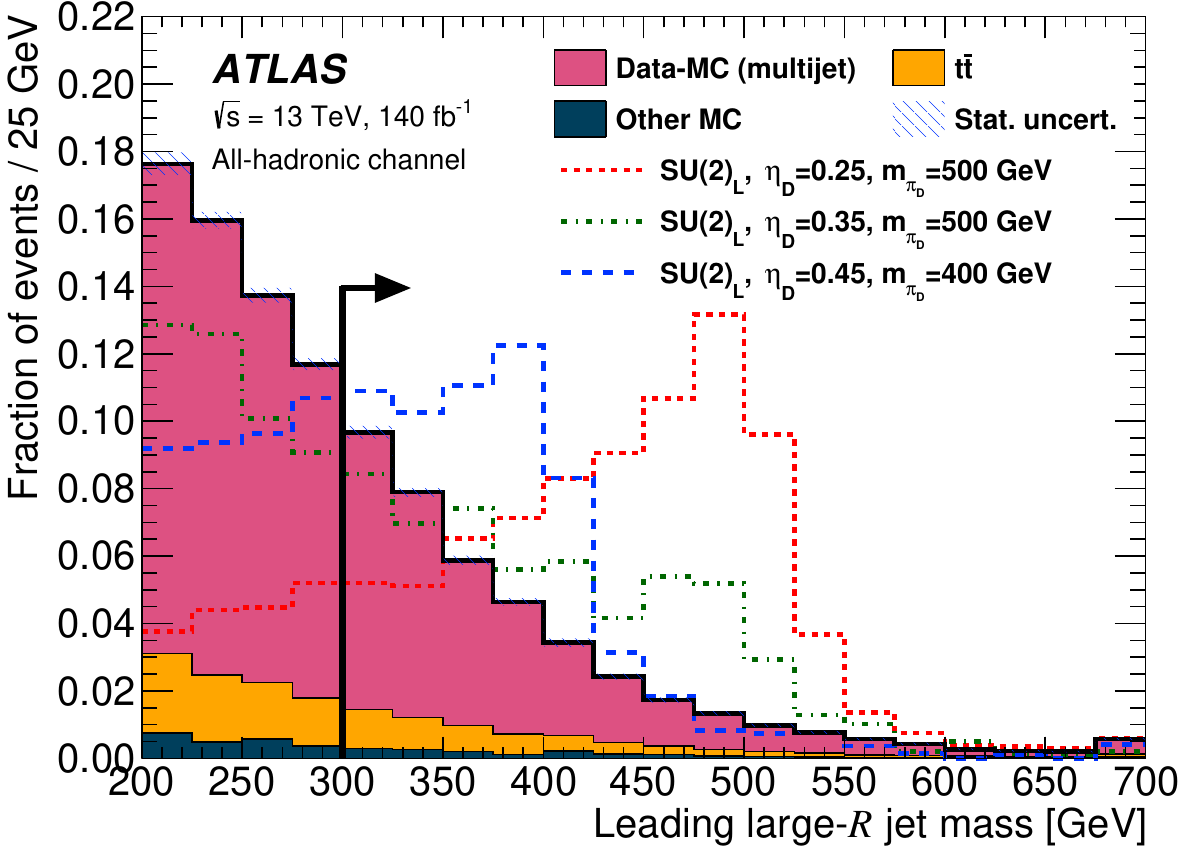}\label{fig:leadingJetMass}}
\subfloat[]{\includegraphics[width=0.495\textwidth]{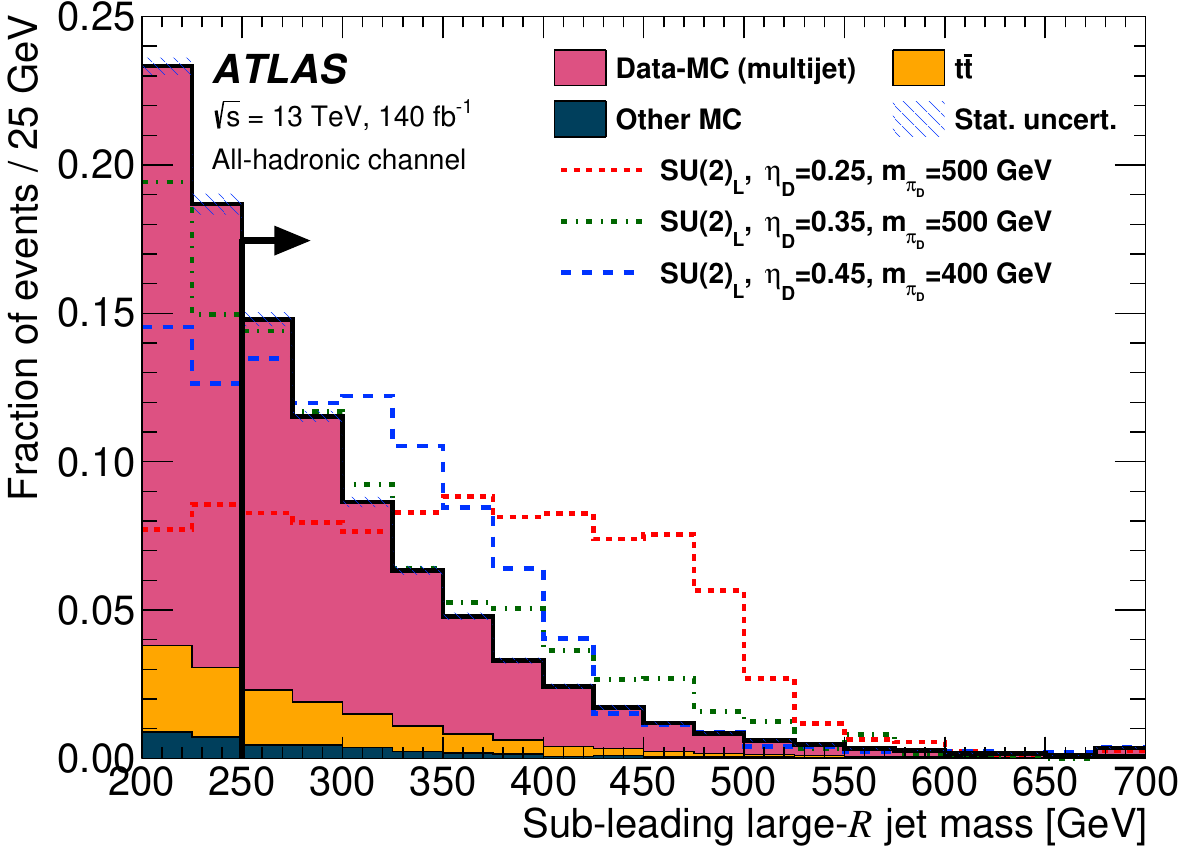}\label{fig:subleadingJetMass}}
\caption{Mass of the \protect\subref{fig:leadingJetMass} leading and \protect\subref{fig:subleadingJetMass} sub-leading large-$R$ jet for all simulated backgrounds overlaid with three example distributions for various signal points after preselection in the \allhad channel. Also shown is a simplified data-driven estimate of the multijet background which was created by taking the event yields for data and subtracting all simulated backgrounds from it. Statistical uncertainties stemming from MC are indicated by the shaded region. The SR is to the right of the vertical line in both the subfigures. Individual SR bins select sub-regions of leading and sub-leading large-$R$ jet mass for improved background discrimination. The last bin contains the overflow.}
\label{fig:rcjet_masses}
\end{figure}
 
\begin{table}[!htb]
\centering
\caption{Summary of selection criteria for the SR (``Tag selection''). Nine bins are defined in the leading large-$R$ jet vs.\ sub-leading large-$R$ jet mass plane. The inverted selection (``Anti-tag selection'') is also defined for use in the data-driven multijet extrapolation described in Section~\ref{sec:bkg_allhad}.}
\begin{tabular*}{\linewidth}{@{\extracolsep{\fill}} l|cc|cc}
\toprule
& Tag & Variable & Tag selection & Anti-tag selection  \\
\toprule
Both large-$R$ jets &  & \mptbb & $> 0.25$ & $> 0.25$ \\
\midrule
Leading large-$R$ jet & $bb_1$ & $\drbj$ &  $<1.0$ &  $\geq 1.0$                           \\
\midrule
Sub-leading large-$R$ jet & $bb_2$ & $\drbj$ &  $<1.0$  &  $\geq 1.0$                     \\
\midrule
\multirow{3}{*}{Leading large-$R$ jet} & \multirow{3}{*}{\pitagone} & \multirow{3}{*}{\makttwelve} & $[300-325~\GeV,$ & \multirow{3}{*}{$\leq 300~\GeV$} \\
& & & $325-400~\GeV,$ & \\
& & & $>400~\GeV]$ & \\
\midrule
\multirow{3}{*}{Sub-leading large-$R$ jet} & \multirow{3}{*}{\pitagtwo} & \multirow{3}{*}{\makttwelve} & $[250-300~\GeV,$ & \multirow{3}{*}{$\leq250~\GeV$ } \\
& & & $300-350~\GeV,$ & \\
& & & $>350~\GeV]$ & \\
\bottomrule
\end{tabular*}
\label{tab:SR_tags}
\end{table}
 
\subsection{One-lepton channel}
\label{sec:evtsel_onelep}
Events satisfying the preselection requirements for the \onelep channel are categorized into SRs, CRs and VRs based on two kinematic variables defined in terms of the properties of the small-$R$ and large-$R$ jets. The first variable, \deltaR, is defined as the angle between the lepton in the event and the second closest $b$-jet to this lepton and aims to suppress \ttbar background similar to $\drbj$ in the all-hadronic channel.
The kinematics of the second \bjet as distinguishing characteristic of signal events is also utilized for the second variable, \bbinvm, defined as the invariant mass of the two $b$-jets in the event that are closest to each other. This is effective for discriminating high dark pion mass signal points against background in which the two $b$-quarks closest to each other come e.g. from gluon splitting.
The distributions of these variables in signal and background MC simulations are shown in Figure~\ref{fig:regiondefinitionvariables} for preselected events.
 
\begin{figure}[btp]
\centering
\subfloat[]{ \includegraphics[width=0.495\textwidth]{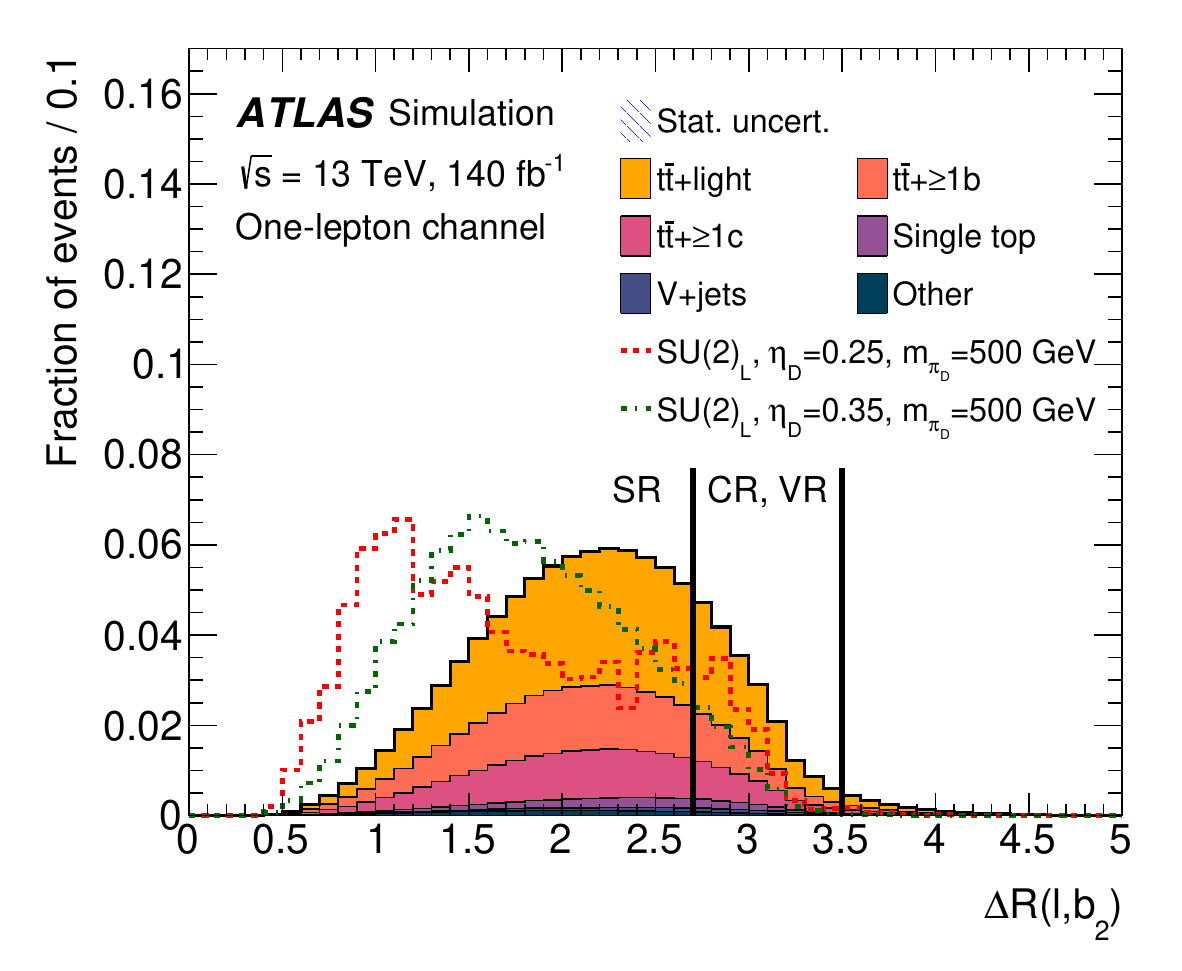}\label{fig:deltaR}}
\subfloat[]{\includegraphics[width=0.495\textwidth]{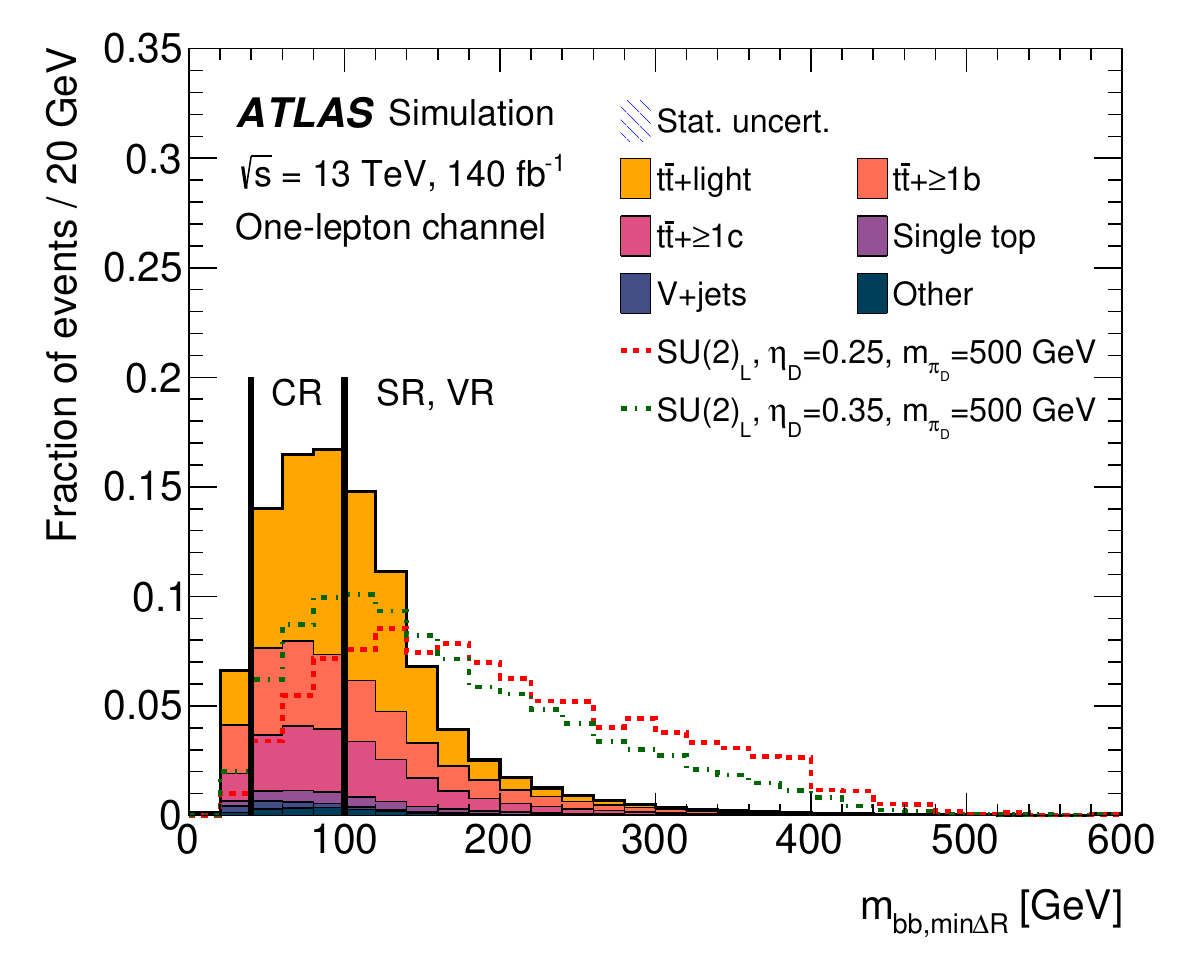}\label{fig:bbm}}
\caption{Normalized distributions of \protect\subref{fig:deltaR} $\deltaR$ and \protect\subref{fig:bbm} \bbinvm for all simulated backgrounds with two example signal distributions overlaid after the \onelep channel preselection. Statistical uncertainties stemming from MC are indicated by a shaded region, but are not visible on the scale of the $y$-axis. The vertical dashed lines indicate the selection requirements applied to events in the SR, CR and VR, as indicated by the labels.}
\label{fig:regiondefinitionvariables}
\end{figure}
 
The SR is defined by requiring $\deltaR < 2.7$ and $\bbinvm > 100~\GeV$.
A CR for the \ttbar+HF background is defined by the requirements {2.7~<~\deltaR~<~3.5}, and {40~\GeV~<~\bbinvm~<~100~\GeV}, thus ensuring orthogonality to the SR. This region is used to correct for mismodeling in \ttbar+HF events and has a background composition similar to that in the SR.
Typical signal contamination in the CR, from signal points that are not already excluded through re-interpretation of other collider searches~\cite{Butterworth:2021jto}, is below 1\%.
The \ttbar+HF background estimate is validated in a VR defined by the requirements {2.7~<~\deltaR~<~3.5}, and {\bbinvm~>~100~\GeV}, making it orthogonal to both the SR and the CR while also exhibiting a background composition similar to that of the SR and CR.

The statistical analysis in the \onelep channel relies on a profile-likelihood fit to the distribution of the sum of the masses of the reclustered jets, described in Section~\ref{sec:reco}, \dave. This discriminating variable is shown in Figure~\ref{fig:dave} for preselected events. For the fit, all events are further classified into regions split into bins based on the number of jets and \bjets in the event. Six bins are defined labeled XR\_5j3b, XR\_5j4b, XR\_6j3b, XR\_6j4b, XR\_7j3b and XR\_7j4b, where the number before the 'j' indicates the number of jets, the number before the 'b' the number of \bjets and 'X' can take on the values 'S' for an SR, 'C' for a CR and 'V' for a VR bin. In all cases, the highest jet or \bjet multiplicity is inclusive, e.g. the region CR\_7j4b is a CR that contains events with $\geq7$ jets and $\geq4$ \bjets.
 
\begin{figure}[btp]
\centering
\includegraphics[width=0.495\textwidth]{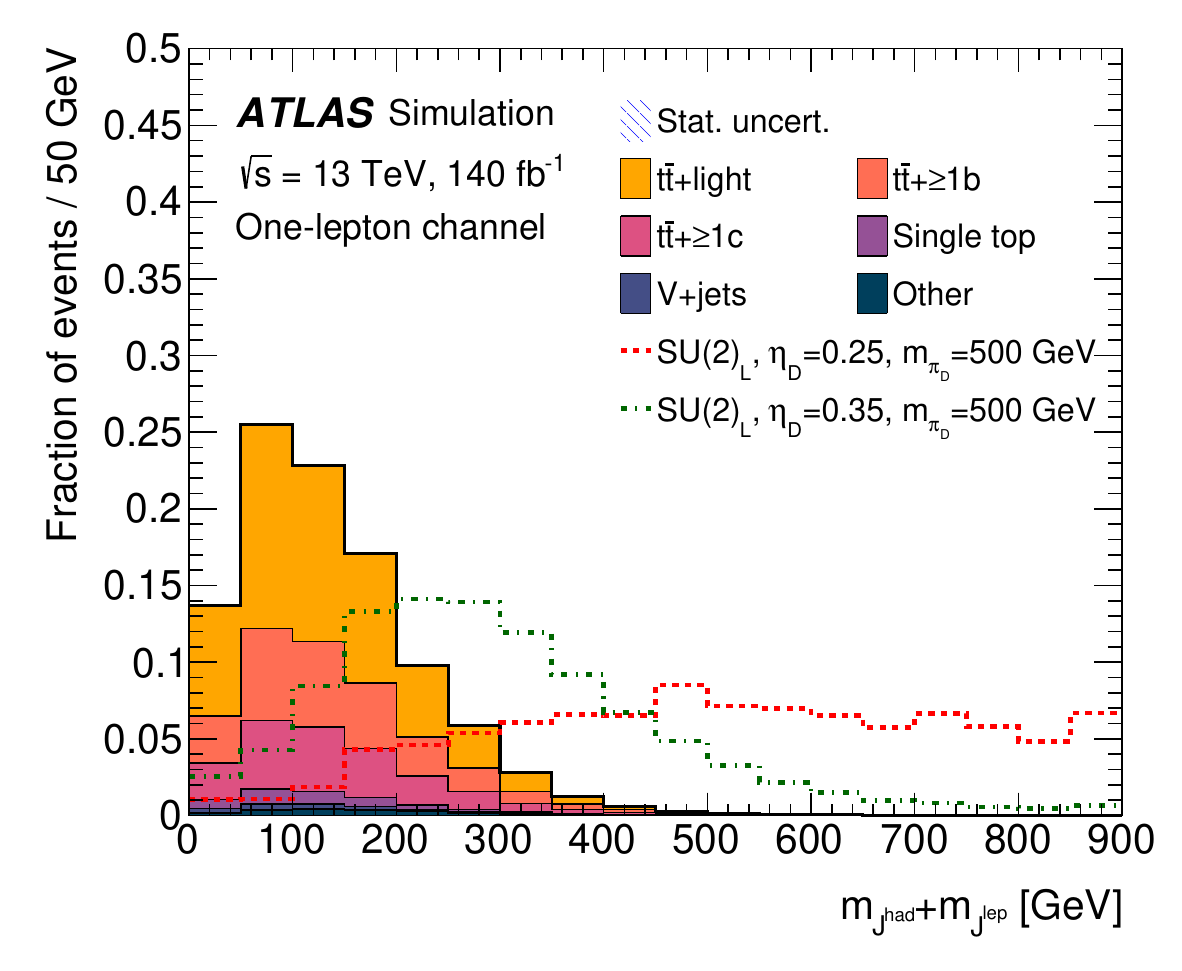}
\caption{Normalized distributions of \dave for all simulated backgrounds overlaid with two example distributions for various signal points after the \onelep channel preselection. Statistical uncertainties stemming from MC are indicated by a shaded region, but are not visible on the scale of the $y$-axis.}
\label{fig:dave}
\end{figure}


\section{Background estimate}
\subsection{All-hadronic channel}
\label{sec:bkg_allhad}

 
The dominant background for the analysis in the \allhad channel originates from multijet events and constitutes about $75\%-85\%$ of the total background in each SR bin. A data-driven method is used to estimate this background, while MC simulation is used to account for the remaining sub-dominant SM processes as described in Section~\ref{sec:samples}. Typical ABCD multijet extrapolations are based on two discriminating variables. Here, however, an extended ABCD method is employed that relies on four instead of two discriminating variables, which allows the correction of correlations between pairs of discriminating variables and provide validation regions close to the SR selection.
 
The multijet background is estimated by extrapolating from regions with small leading and sub-leading \makttwelve and large \drbj to SR bins with large leading and sub-leading \makttwelve and small \drbj. The method is similar to the one detailed in Ref.~\cite{TOPQ-2016-09}. To this end, two additional anti-tags denoted by a slashed tag label (with orthogonal selections to the already described $bb_i$ and $\pitagi$ tags) are defined. The tag $\cancel{bb_i}$ inverts the $bb_i$ selection, while $\cancel{\pitagi}$ places upper requirements on the large-$R$ jet mass, as summarized in Table~\ref{tab:SR_tags}. The combinations of possible tags and anti-tags in an event result in 16 separate regions shown in Figure~\ref{fig:ABCDlabels}. The extended ABCD method extrapolates from regions with one tag to each of the nine SR bins. Two-tag regions are used to determine correlation correction factors and three-tag regions are used for validation of the multijet estimate.
 
\begin{figure}[tb]
\centering
\includegraphics[width=0.65\textwidth]{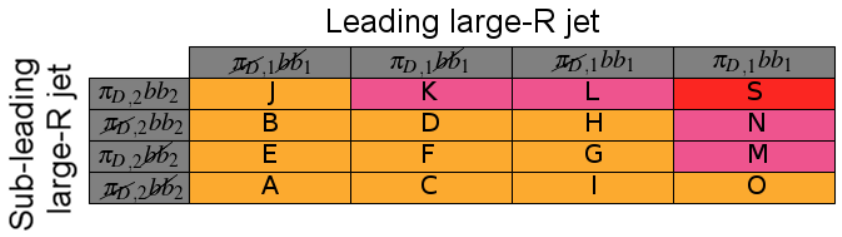}
\caption{Region labels for the 16 regions used in the data-driven multijet estimate in the \allhad channel. Region $S$ labels the SR, regions $B$, $C$, $E$ and $I$ are used for the ABCD extrapolation, regions $D$, $F$, $G$, $H$, $J$ and $O$ are used to compute correlation correction factors, and regions $K$, $L$, $M$ and $N$ are validation regions. The background estimate is performed independently for all nine SR bins.}
\label{fig:ABCDlabels}
\end{figure}
 
The concept of the extended ABCD method is detailed here by explicitly stating the computations for one VR, however, analogous derivations have to be carried out for all SRs and VRs. Considering only region $K$ in Figure~\ref{fig:ABCDlabels}, a 2-variable ABCD estimate for this region would be computed in the standard way through $\hat{K} = \frac{J\cdot D}{B}$, where $J$, $D$ and $B$ are the number of data events minus the number of simulated events in the respective regions. If $\pi_{D,1}$ and $\pi_{D,2}$ are uncorrelated, $\hat{K}$ would be a valid estimate. However, if $\pi_{D,1}$ and $\pi_{D,2}$ are correlated, then $\hat{K}$ needs to be corrected by a correlation factor ($k$-factor). As long as there is no significant additional three-tag correlation with the $bb_2$ tag, this correlation factor, $k_{\pi_{D,1},\pi_{D,2}}$, can be measured from $\frac{F\cdot A}{C\cdot E}$ since $\frac{K\cdot B}{J\cdot D}=\frac{F\cdot A}{C\cdot E}$. Thus,
\begin{align}
\hat{K} &= \frac{J\cdot D}{B} \cdot \frac{F\cdot A}{C\cdot E} = \frac{J\cdot D}{B} \cdot k_{\pitagone,\pitagtwo}~.
\label{eq:K2var}
\end{align}
 
One can also consider the estimate of region $K$ using a three-variable ABCD estimate computed with two correlation correction factors:
\begin{align}
\hat{K} &= \frac{J\cdot C}{A} \cdot k_{\pitagone,\pitagtwo} \cdot k_{\pitagone,bb_2}~,
\label{eq:K3var}
\end{align}
where $k_{\pitagone,bb_2}=\frac{D\cdot A}{B\cdot C}$.
Substituting this correlation factor into Eq.~(\ref{eq:K3var}) yields once again Eq.~(\ref{eq:K2var}).
All other $k$-factors can be defined according to the same principle as $k_{\pi_{D,1},\pi_{D,2}}$ and $k_{\pitagone,bb_2}$.
 
The final multijet background estimate requires a four-variable ABCD estimate $\hat{S}'$ that is computed from data minus event counts from MC simulated backgrounds in the regions $B$, $C$, $E$, and $I$ where exactly one tag is applied and region $A$ with no applied tags according to
\begin{align}
\hat{S}' &= \frac{B\cdot C\cdot E\cdot I}{A^3}~.
\label{eq:ABCD_Sprime}
\end{align}
This estimate is then multiplied by six $k$-factors to correct for correlations between tags,
\begin{align}
\label{eq:ABCD_S}
\hat{S} &= \hat{S}' \cdot k_{\pitagone,bb_1} \cdot k_{\pitagtwo,bb_2} \cdot k_{\pitagone,bb_2} \cdot k_{\pitagtwo,bb_1} \cdot k_{\pitagone,\pitagtwo} \cdot k_{bb_1,bb_2}~.
\end{align}
 
If the selection criteria defined by these tags are independent from each other then the expectation value of the corresponding $k$-factor will be $1$. Correlation factors around $1.5$ are observed between the \pitagone and $bb_1$ tags as well as the \pitagtwo and $bb_2$ tags. The $bb_1$ and $bb_2$ tags are highly correlated with a $k$-factor of $0.1$ due to the preselection requirement of three $b$-tagged jets. This could be mitigated by requiring four $b$-tagged jets, however, this would result in low statistics and high signal contamination for the four-variable ABCD extrapolation regions.
 
Typical signal contaminations in the extrapolation regions are less than $5\%$ and small compared to the uncertainty applied to the multijet background discussed in Section~\ref{sec:systematics}. The method is validated through the closure of the estimate in the four 3-tag validation regions $K$, $L$, $M$ and $N$. Figure~\ref{fig:VR_hists} compares data to estimated background yields for each validation region in each bin of the SR. If significant 3-tag correlations occurred, a discrepancy between multijet estimate and data yields should be visible in the validation regions. However, all data yields are compatible with the background estimate within the uncertainty bands for all validation regions. The uncertainty includes a non-closure systematic uncertainty, which by design covers non-linear correlations and the impact of multijet estimation regions with low statistics. It ranges between $33\%$ and $57\%$ and is derived and discussed in Section~\ref{sec:systematics}. The procedure was further validated by performing signal injection tests as well as stability tests over time, under variation of the selection criteria and with scaled simulated background contributions. In all cases the resulting background estimates where stable and thus consistent with the nominal estimate.
 
\begin{figure}[tb]
\centering
\includegraphics[width=0.99\textwidth]{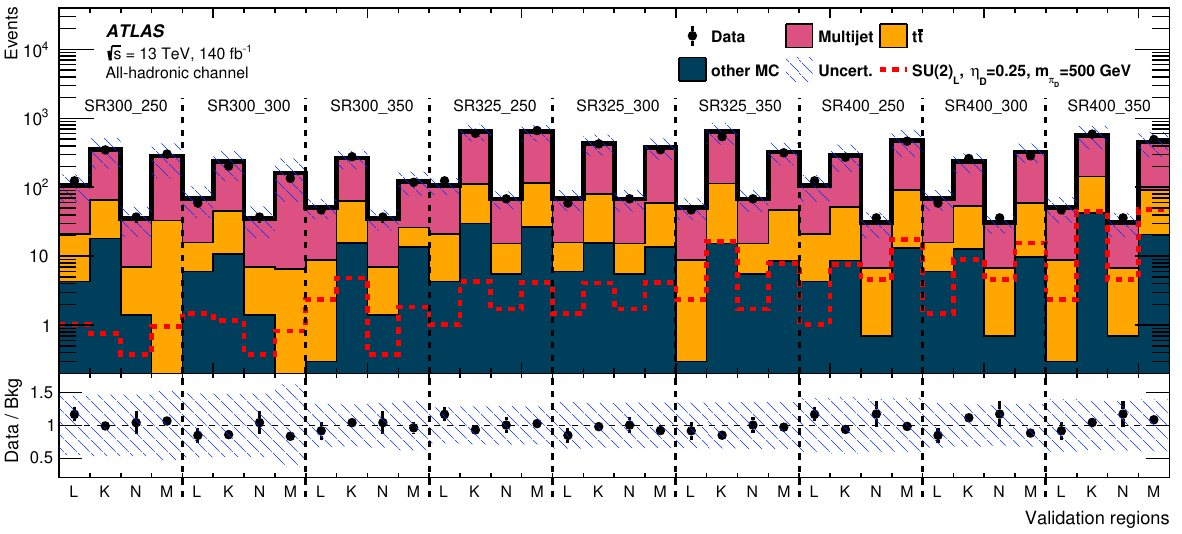}
\caption{The four \allhad channel validation regions $K$, $L$, $M$ and $N$ from Figure~\ref{fig:ABCDlabels} for each of the nine SR mass bins with the following naming convention: the leading large-$R$ jet lower mass boundary in \GeV is followed by the sub-leading large-$R$ jet lower mass boundary. The shaded region indicates the uncertainty on the background estimate in each bin that includes the statistical uncertainties from the limited sample sizes in data and simulation as well as a multijet non-closure systematic uncertainty, as detailed in Section~\ref{sec:systematics}.
}
\label{fig:VR_hists}
\end{figure}


\subsection{One-lepton channel}
\label{sec:bkg_onelep}

The background in the \onelep channel is estimated from MC simulations using the samples described in Section~\ref{sec:samples} and is dominated by \ttbar production in association with light-flavor or heavy-flavor quarks. This background is estimated by using the inclusive \ttbar sample complemented by the dedicated \ttbb sample, with events categorized into \ttb, \ttc and \ttlight as described in Section~\ref{sec:samples}. The \ttbar+HF background that populates the SR is known from previous studies to be underestimated by the current MC predictions~\cite{TOPQ-2017-12}. This mismodeling is corrected by keeping the normalization of these backgrounds unconstrained in the profile-likelihood fit to the \dave distribution. To this end, the \ttc and \ttlight contributions are combined and two normalization factors are fit depending on whether in addition to \ttbar any $b$-quarks are present in the event. In this categorization, the combination of the \ttc and \ttlight backgrounds is referred to as \ttnb. The resulting background estimates for the SR, CR and VR are presented in Section~\ref{sec:fit_onelep}.


\section{Systematic uncertainties}
\label{sec:systematics}

 
The predicted signal and background event yields in the SR bins are affected by various sources of systematic uncertainties stemming from instrumentation, the data-driven multijet estimation and theoretical considerations.
For the \allhad channel, the total uncertainty is dominated by the uncertainty on the multijet background estimate.
In the \onelep channel, the theoretical sources of uncertainty on the background modeling dominate the total uncertainty.
 
\subsection{Experimental uncertainties}
The uncertainty in the combined 2015--2018 integrated luminosity is $0.83\%$~\cite{DAPR-2021-01}, obtained using the LUCID-2 detector~\cite{LUCID2} for the primary luminosity measurements, complemented by measurements using the ID and calorimeters. A systematic variation that might be introduced by the reweighting of simulated samples to match the pileup profile observed in data is estimated by varying the scale factor applied to the pileup distributions.
The onset of the \Ht trigger used in the \allhad channel was studied and potential effects of mismodeling were evaluated. All three of these uncertainties were found to have a negligible impact on the analysis ($<1\%$ in the \allhad channel, and ranking from $<1\%$ to a few percent depending on the signal in the \onelep channel).
 
Slight performance differences between data and simulation in lepton reconstruction, identification and isolation are corrected by the application of scale factors that are estimated from tag-and-probe experiments in data and simulation~\cite{EGAM-2018-01,MUON-2018-03}. The impact of lepton momentum scale corrections is evaluated by $\pm1\sigma$ scale variations. For resolution uncertainties the lepton energy or momentum is smeared. In total seven (twelve) separate variations for electrons (muons) are considered. The impact of lepton uncertainties is less than $1\%$ in both channels.
 
The determination of the jet energy scale and resolution is done by combining information from collision data, test beam data and simulation as described in Ref.~\cite{JETM-2018-05}. Effects from jet flavor composition, single-particle response and pileup are considered. In the \onelep channel, 29 parameters are evaluated for scale variations, while 13 parameters are evaluated for jet \pT\ resolution systematic uncertainties.
In the \allhad channel the variations are simplified since the data-driven background estimation method largely compensates yield changes by different systematic variations and causes most of the systematic uncertainties to be negligibly small. As such, 23 parameters arise from scale variations, while for jet \pT\ resolution systematic uncertainties 8 parameters are evaluated. The impact of these uncertainties on the final result is small, ranging between $<1\%$ to about $4\%$ in the \allhad channel, and remaining $<10\%$ in the \onelep channel.
 
Uncertainties on the corrections of $b$-tagging efficiency differences between data and simulation are derived from dedicated flavor-enriched subsets of the data~\cite{FTAG-2018-01}. Also considered are uncertainties due to the mis-tagging of $c$-jets~\cite{FTAG-2020-08} and light-flavor jets~\cite{FTAG-2019-02}. 
Additionally, variations to extrapolate the measured uncertainties to the high-\pt region are considered for both channels~\cite{ATL-PHYS-PUB-2021-003}. The impact of flavor-tagging systematic uncertainties is less than $1\%$ on the final result in the \allhad channel, while a larger contribution is observed in the \onelep channel, reaching values of the order of 10\%.
 
\subsection{Modeling uncertainties in background simulations}
For the \allhad channel, uncertainties in modeling of the \ttbar background are included. 
For the \onelep channel, uncertainties in modeling the \ttbar, \ttbb, and single top-quark backgrounds are included. All other simulated backgrounds are negligible in both channels and no systematic uncertainties are assigned to them. For the single top-quark background, a 30\% normalization uncertainty is applied~\cite{EXOT-2022-14} with an impact ranging between $<1\%$ and a few percent. Details on the \ttbar and \ttbb uncertainties are described below.
 
\subsubsection{\ttbar uncertainties}
\label{sec:ttbaruncertainty}
Several uncertainties in the theoretical modeling of the \ttbar background samples are considered. In the \onelep channel the \ttbar theory systematic uncertainties apply only on the \ttlight and \ttc background components as they are estimated from the five-flavor scheme \ttbar sample, while the \ttb background has dedicated systematic uncertainties.
 
Missing higher order contributions in perturbative expansion of the \ttbar production cross-section are estimated by adding in quadrature contributions from renormalization and factorization scale variations, which are obtained by independently varying the parameters $\mu_R$ and $\mu_F$ by a factor $0.5$ and $2.0$ and taking the envelope.
Uncertainties on the choice of PDF set used for event simulation are estimated by using the PDF4LHC and NNPDF error sets following the PDF4LHC prescription~\cite{ThePDF4LHCWorGroIntRec} and taking the envelope.
The initial-state radiation (ISR) modeling is estimated by variations of the strong coupling constant $\alpha_S$ through the \textsc{Var3c} tune variation.
The amount of final-state radiation (FSR) in an event is estimated by varying the factorization scale by factors $0.5$ and $2.0$ inside \pythia.
To assess the uncertainty in the matching of NLO matrix elements to the parton shower, the nominal sample is compared to an alternative sample obtained setting the \texttt{pthard} \pythia parameter to 1 (the default is 0). This parameter regulates the definition of the vetoed region of the showering, important to avoid holes and overlaps in the phase space filled by \powheg and \pythia. This recommendation follows the description included in Ref.~\cite{Hoche:2021mkv}. The alternative sample was produced using \powheg interfaced with \pythia.306 using the \nnpdftwo PDF set and the A14 set of tuned parameters.
The impacts of using a different parton shower and hadronization model were evaluated by comparing the nominal \ttbar sample with another event sample produced with the \powhegbox~v2 generator. For the parton shower variation the \nnpdfnlo PDF set was used, while events in the sample used to estimate the impact of the hadronization model were interfaced with \herwig~\textsc{7.04}~\cite{Bahr:2008pv,Bellm:2015jjp}, using the H7UE set of tuned parameters~\cite{Bellm:2015jjp} and the \mmhtlo PDF set~\cite{Harland-Lang:2014zoa}. The impact of a variation of the \hdamp parameter is assessed by comparing the nominal samples to an alternative set of samples for which \hdamp is set to $3m_{\text{top}}$.
 
All alternative samples used to derive the systematic uncertainties are corrected to match the NNLO in QCD and NLO in EW predictions of the top/antitop-quark \pt and the top-quark mass distribution using the procedure outlined in Section~\ref{sec:ttbarBackground}.
A systematic uncertainty associated with this reweighting itself is derived from the maximum and minimum 7-point scale variations, independently for the top/antitop-quark \pt and the top-quark mass. The variations are taken into account in the final statistical fit by including them as scale variations on the \ttbar background.
 
To avoid over-constraining the \ttbar modeling nuisance parameters in the \onelep channel, the theoretical systematic uncertainties are treated as uncorrelated among \ttlight and \ttc, jet and \bjet multiplicity bins, and between their shape and acceptance components.
 
The impact of \ttbar modeling uncertainties on the final result is found to range between $<1\%$ and $10\%$.
 
\subsubsection{\ttbb uncertainties}
\label{sec:syst_ttbb_theory}
 
Theory uncertainties on the \ttbb sample are only applied to the \ttb component in the \onelep channel, as all other \ttbar components are estimated from the bulk \ttbar sample.
 
The scale, PDF, ISR, and FSR uncertainties for the \ttbb sample are derived in the same way as the bulk \ttbar sample. The impacts of using a different parton shower and hadronization model are evaluated by comparing the nominal \ttbar sample to another sample produced with the \powhegbox~v2 generator. For the parton shower variation the \nnpdfnlo PDF set was used, while events in the sample estimating the impact of the hadronization model were interfaced with \herwig~\textsc{7.1}, using the \textsc{H7.1-Default} set of tuned parameters and the \mmhtlo PDF set~\cite{Harland-Lang:2014zoa}.
 
The matching uncertainty is evaluated by comparing the nominal sample with an alternative sample obtained by setting the \texttt{pthard} \pythia parameter to 1. The alternative sample was produced using \powheg interfaced with \pythia.307 using the \nnpdftwo PDF set and the A14 set of tuned parameters.
 
The \ttbb uncertainty nuisance parameters are treated as uncorrelated between the different regions of jet and \bjet multiplicities and are further split up into their shape and acceptance components to avoid over-constraining them in the fit. They constitute the dominant systematic contribution in the \onelep channel, with an impact on the final results ranging from 1\% to about 30\%.

\subsection{Data-driven background estimation uncertainties}
 
In the \allhad channel, the dominant systematic uncertainty originates from the multijet estimation method. Deviations of the ratio between data and SM estimate from unity in the VRs, denoted by $k_{VR} = (\text{data}-\text{MC})/\text{multijet}$, are used to estimate a non-closure systematic uncertainty. These $k$-factors are a measure of the remaining correlations between 3-tags in the multijet estimation method and are expected to be close to unity. The statistical uncertainties on the $k$-factors $\sigma_{k_{VR}}$ are added in quadrature so that the final multijet uncertainty is calculated according to:
\begin{equation}
\sigma_{ABCD} = \sqrt{ \left(1 - \prod_{VR} k_{VR}\right)^2+\sum_{VR}\sigma^2_{k_{VR}}}\quad,
\end{equation}
where the first term under the square root describes the non-closure component of the systematic uncertainty calculated from the $k$-factors and the second term the statistical uncertainty summed up in quadrature over all four VRs. The statistical component dominates in most regions and is driven by region $H$ in Figure~\ref{fig:ABCDlabels} which has the lowest number of events. The computation is done separately for each bin of the SR producing values ranging from $33\%$ to $57\%$. The uncertainty in each SR mass bin is treated as fully uncorrelated with the other SR mass bins thus yielding a conservative estimate.


\section{Statistical analysis and results}
\label{sec:fit}

Signal hypotheses are evaluated against the data by performing profile-likelihood fits in each analysis channel.
The set of nuisance parameters $\theta$ that corresponds to the systematic uncertainties is scaled in such a fashion that, before the fit, all individual uncertainties $j$ have $\theta_j=0$. In addition, the uncertainties of these nuisance parameters $\sigma$ are scaled to fulfill $\sigma_j=1$ before fitting. Defining $\gamma$ as the set of statistical uncertainties with $\gamma_i$ the uncertainty on the number of predicted events of a specific bin $i$, the set is scaled to be a factor around 1. 
The profile-likelihood fit is then performed by minimizing the quantity $q(\mu)=-2\log\mathcal{L}$, where the likelihood $\mathcal{L}$ is defined as
\begin{equation}
\mathcal{L}(\mu,\theta|S,B,N)=\prod_{i\in\text{bins}}\mathcal{P}\left(N_i|\mu S_i+B_i\right)\times\mathcal{P}\left(S_i+B_i|\gamma_i\right)\prod_{j\in\text{syst.}}\mathcal{G}\left(\theta_j,\sigma_j\right)
\label{eq:likelihood}
\end{equation}
with $N$ being the number of data events, $S$ the predicted number of signal events and $B$ the expected number of background events.
Statistical and systematic uncertainties are taken to be Poissonian ($\mathcal{P}$) and Gaussian ($\mathcal{G}$) distributed parameters, respectively.
The fit is evaluated using the \textsc{RooFit}~\cite{RooFit} package with the minimization conducted with the \textsc{Minuit2} package~\cite{ref:1589282}, which yields the optimal values for $\mu$ and the set of parameters $\theta_i$, $\gamma_i$ and $\sigma_i$ in a signal plus background fit. 
 
\subsection{All-hadronic channel}
\label{sec:fit_allhad}
 
The fit in the \allhad channel is performed over all nine SR bins simultaneously.
Since the multijet estimate constrains background to the data, most systematic uncertainties (other than from the multijet estimate) have limited impact on the final result and all systematic uncertainties below $1\%$ are ignored. Surviving uncertainties are kept as nuisance parameters in the fit. Because simulated backgrounds constitute less than $20\%$ of the total SM background in the SR, no CRs are used to scale the simulated estimates.
 
\begin{figure}[tbp]
\centering
\includegraphics[width=0.495\textwidth]{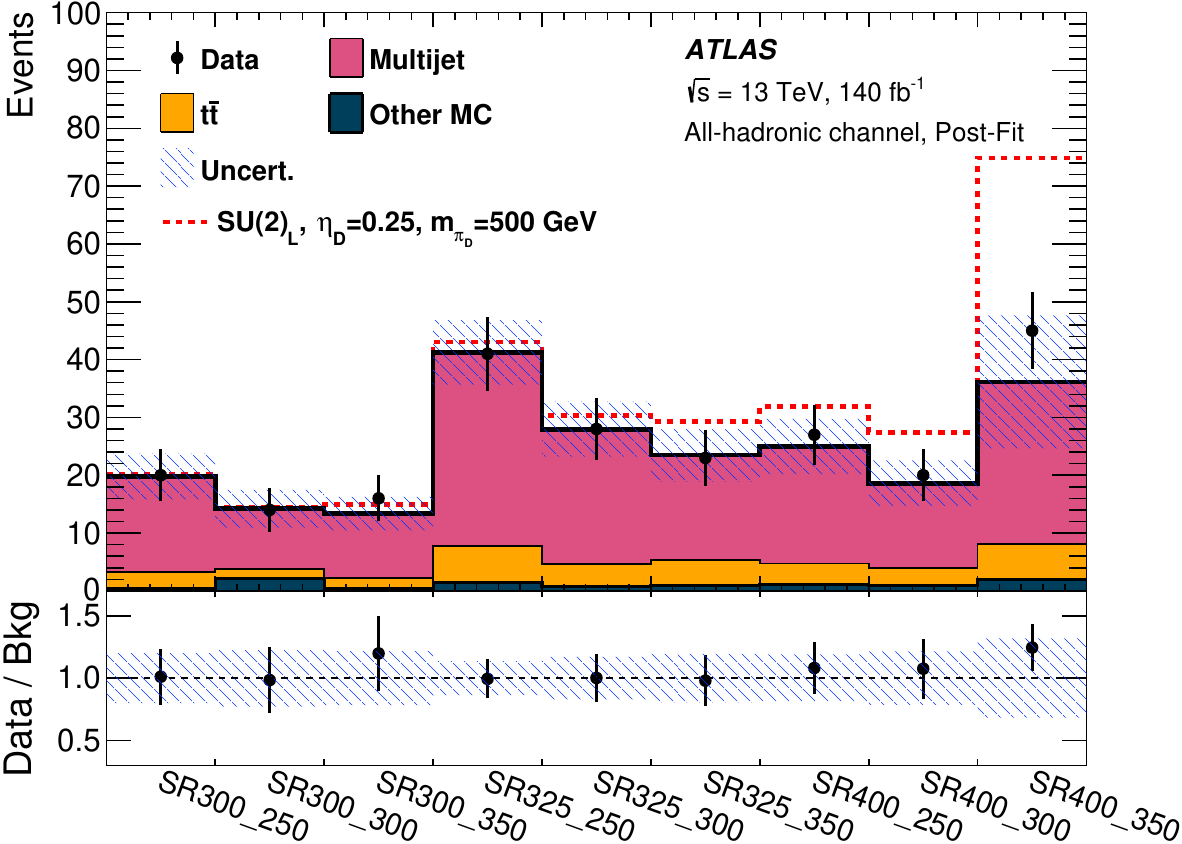}
\caption{Comparison of the events yields in data and the background prediction after the fit under the background-only hypothesis in all bins of the \allhad SR. The dashed line is stacked onto the SM background and shows an example \sul signal with $\etaD=0.25$ and $\dpmass=500~\GeV$. The uncertainty bands contain all statistical and systematic uncertainties. The horizontal axis labels reflect the different SR bins, with the first number indicating the lower boundary of the leading large-$R$ jet mass bin and the second number the lower boundary of the sub-leading large-$R$ jet mass bin.}
\label{fig:postfit_distribution}
\end{figure}
 
Figure~\ref{fig:postfit_distribution} shows the distributions in all SR bins after fitting. No significant pulls or constraints are observed for the fitted nuisance parameters. The results are also summarized in Table~\ref{tab:allhad_sr_yields}. As can be seen the data are very well described by the background-only hypothesis with no significant excess of events above the SM prediction.
 
\begin{table}[htb]
\caption{Observed and predicted event yields after the fit under the background-only hypothesis in all nine \allhad SR mass bins. The name of each column corresponds to the SR bins: the leading large-$R$ jet lower mass boundary in \GeV is followed by the sub-leading large-$R$ jet lower mass boundary; the same convention is used in Figure~\ref{fig:postfit_distribution}. The quoted uncertainties contain statistical and systematic components. The total post-fit uncertainty can be smaller than the sum in quadrature of the different components due to correlations resulting from the fit to data.}
\label{tab:allhad_sr_yields}
\centering
\begin{tabular}{c|r@{}lr@{}lr@{}l}
& \multicolumn{2}{c}{SR300\_250} & \multicolumn{2}{c}{SR300\_300} & \multicolumn{2}{c}{SR300\_350} \\
\toprule
V+jets & \multicolumn{2}{c}{$<0.01$}  & $\,\,1.97\,\pm\,$ & $0.89$ & $\,\,0.28\,\pm\,$ & $0.06$ \\
Single top & $0.12\,\pm\,$ & $0.07$ & $0.00\,\pm\,$ & $0.03$ & \multicolumn{2}{c}{$<0.01$}  \\
$t\bar{t}+X$ & $0.30\,\pm\,$ & $0.04$ & $0.21\,\pm\,$ & $0.09$ & $0.17\,\pm\,$ & $0.04$  \\
$t\bar{t}$ & $2.9\,\pm\,$ & $1.9$ & $1.6\,\pm\,$ & $1.1$ & $1.78\,\pm\,$ & $0.76$ \\
Multijet & $16.5\,\pm\,$ & $4.3$ & $10.5\,\pm\,$ & $3.6$ & $11.1\,\pm\,$ & $3.1$  \\
\midrule
Total SM & $19.8\,\pm\,$ & $4.0$ & $14.2\,\pm\,$ & $3.3$ & $13.4\,\pm\,$ & $2.9$ \\
\midrule
Data & \multicolumn{2}{c}{20} & \multicolumn{2}{c}{14} & \multicolumn{2}{c}{16} \\
\bottomrule
\multicolumn{4}{c}{\phantom{0}}\\
 
& \multicolumn{2}{c}{SR325\_250} & \multicolumn{2}{c}{SR325\_300} & \multicolumn{2}{c}{SR325\_350} \\
\toprule
V+jets & $\,\,0.74\,\pm\,$ & $0.64$ & $\,\,0.12\,\pm\,$ & $0.18$ & $\,\,0.19\,\pm\,$ & $0.16$ \\
Single top & $0.36\,\pm\,$ & $0.06$ & $0.12\,\pm\,$ & $0.13$ & $0.27\,\pm\,$ & $0.15$ \\
$t\bar{t}+X$ & $0.35\,\pm\,$ & $0.06$ & $0.44\,\pm\,$ & $0.11$ & $0.50\,\pm\,$ & $0.07$ \\
$t\bar{t}$ & $6.4\,\pm\,$ & $3.6$ & $4.0\,\pm\,$ & $2.0$ & $4.4\,\pm\,$ & $2.1$ \\
Multijet & $33.4\,\pm\,$ & $6.7$ & $23.3\,\pm\,$ & $5.2$ & $18.1\,\pm\,$ & $5.0$ \\
\midrule
Total SM & $41.2\,\pm\,$ & $5.6$ & $28.0\,\pm\,$ & $4.8$ & $23.5\,\pm\,$ & $4.6$ \\
\midrule
Data & \multicolumn{2}{c}{41} & \multicolumn{2}{c}{28} & \multicolumn{2}{c}{23} \\
\bottomrule
\multicolumn{4}{c}{\phantom{0}}\\
 
& \multicolumn{2}{c}{SR400\_250} & \multicolumn{2}{c}{SR400\_300} & \multicolumn{2}{c}{SR400\_350} \\
\toprule
V+jets & $\,\,0.71\,\pm\,$ & $0.55$ & \multicolumn{2}{c}{$<0.01$}  & $\,\,1.19\,\pm\,$ & $0.30$ \\
Single top & \multicolumn{2}{c}{$<0.01$}  & $0.47\,\pm\,$ & $0.08$ & $0.11\,\pm\,$ & $0.02$ \\
$t\bar{t}+X$ & $0.34\,\pm\,$ & $0.07$ & $0.40\,\pm\,$ & $0.07$ & $0.73\,\pm\,$ & $0.10$ \\
$t\bar{t}$ & $3.7\,\pm\,$ & $1.8$ & $3.1\,\pm\,$ & $2.5$ & $6.1\,\pm\,$ & $3.9$ \\
Multijet & $20.2\,\pm\,$ & $5.3$ & $14.7\,\pm\,$ & $4.6$ & $28\,\pm\,$ & $12$ \\
\midrule
Total SM & $25.0\,\pm\,$ & $4.8$ & $18.6\,\pm\,$ & $4.0$ & $36\,\pm\,$ & $11$ \\
\midrule
Data & \multicolumn{2}{c}{27} & \multicolumn{2}{c}{20} & \multicolumn{2}{c}{45} \\
\bottomrule
\end{tabular}
\end{table}
 
\subsection{One-lepton channel}
\label{sec:fit_onelep}
In the \onelep channel, a profile-likelihood fit is performed to the binned \dave spectrum in the six SRs and the six CRs defined based on jet and \bjet multiplicity requirements, simultaneously.
As already indicated in Section~\ref{sec:bkg_onelep}, two independent normalization factors are used for the \ttbar background depending on whether in addition to \ttbar any $b$-quarks are present in the event. These two normalization factors are denoted by \kttb and \kttnb.
To avoid over-constraining the background modeling nuisance parameters, the theoretical systematic uncertainties are treated as uncorrelated among different \ttbar flavors, jet and \bjet multiplicity bins, and between their shape and acceptance components.
 
The event yields for all SRs, CRs and VRs are shown after a fit to the data under the background-only hypothesis in Figure~\ref{fig:1lep_summary}.
The corresponding distributions of \dave are shown in Figure~\ref{fig:1lep_postfit_SR} for all jet and \bjet multiplicity bins of the SR. The SM background estimate is in good agreement with data after the background-only fit. The post-fit normalization factors for the \ttbar background components are $\kttnb=0.92^{+0.09}_{-0.08}$ and $\kttb=1.60^{+0.19}_{-0.17}$. No pulls greater than 1$\sigma$ are observed for the fitted nuisance parameters, with only some constraints in the nuisance parameters describing the theoretical modeling. Data and post-fit background event yields are shown in Table~\ref{tab:1lep_sr_yields}. Good agreement between the background prediction and data is observed in all regions. 
 
\begin{figure}[btp]
\centering
\includegraphics[width=0.995\textwidth]{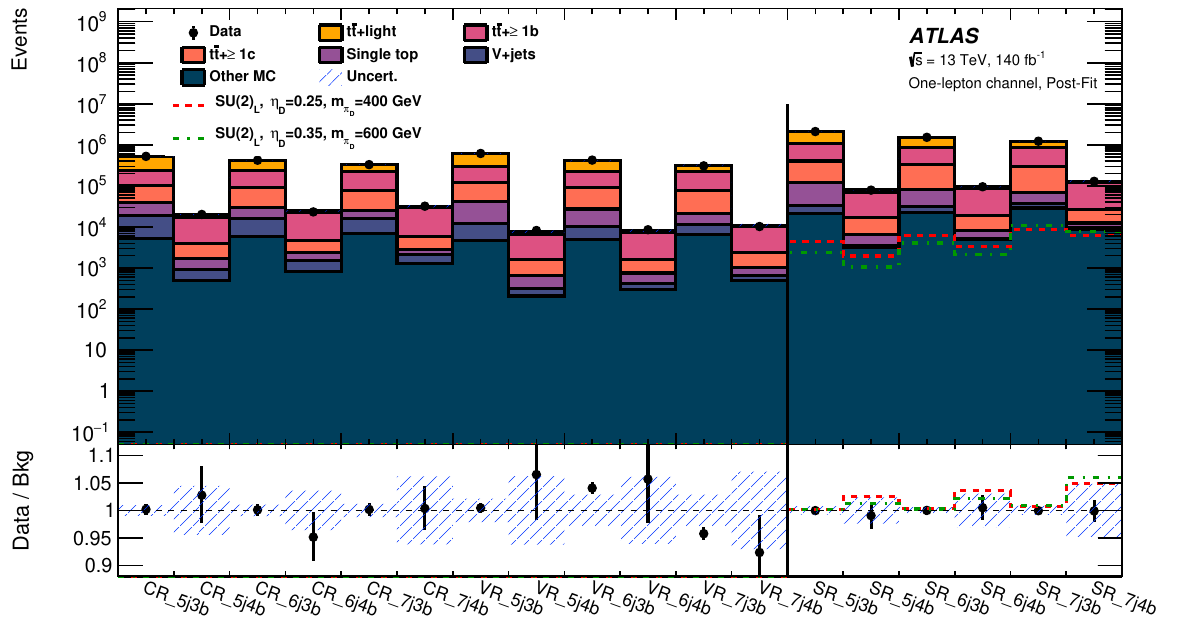}\label{fig:1leppostfit}
\caption{Comparison of the events yields in data and the background prediction in the SRs, CRs, and VRs in the \onelep channel, after the fit to data in all CRs and SRs under the background-only hypothesis. The uncertainty bands contain all statistical and systematic uncertainties. Two example \sul signal points are overlaid in the SRs.}
\label{fig:1lep_summary}
\end{figure}
 
\begin{figure}[btp]
\centering
\subfloat[]{\includegraphics[width=0.35\textwidth,page=7]{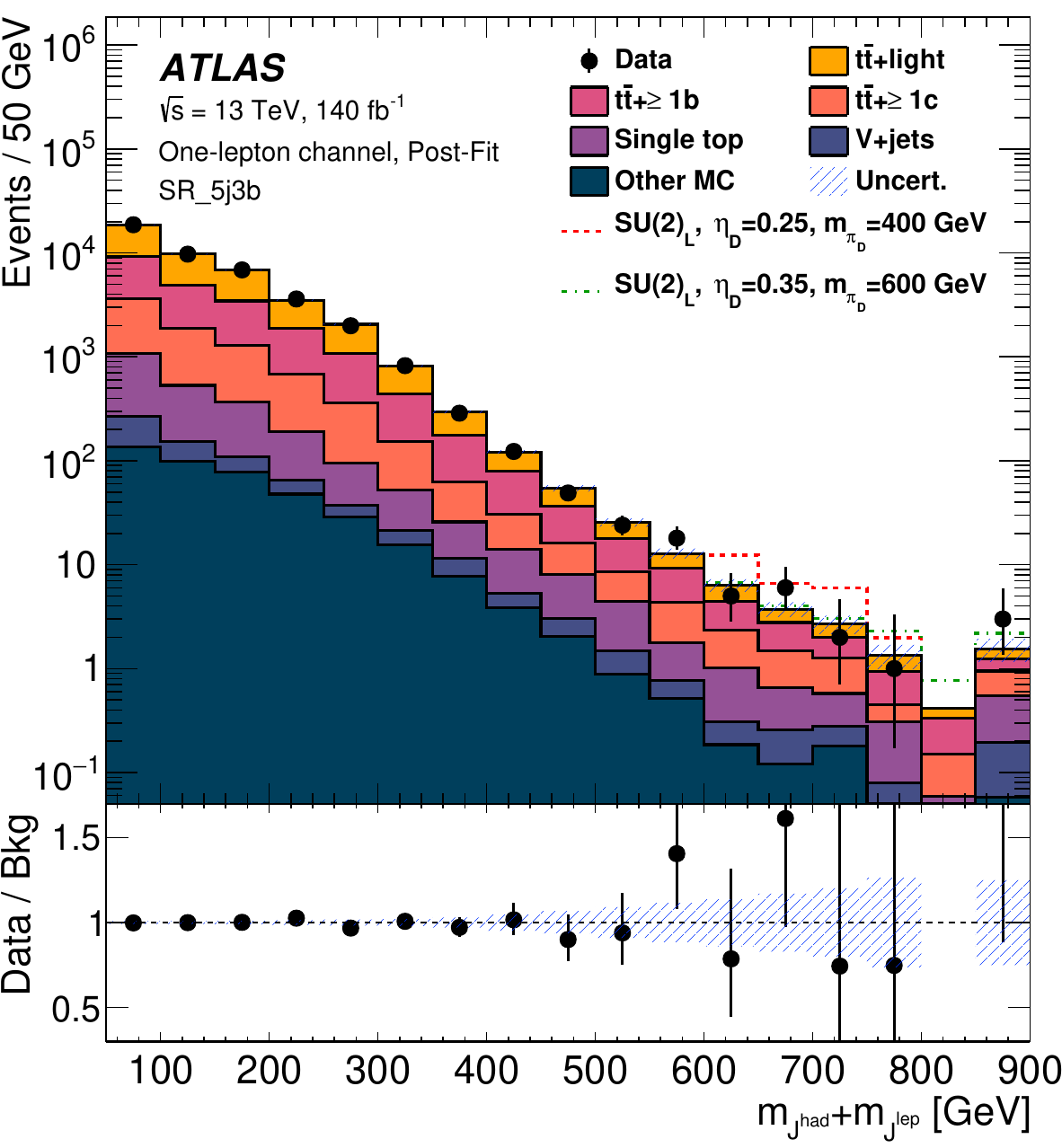}\label{fig:1lep_postfit_SR5j3b}}
\subfloat[]{\includegraphics[width=0.35\textwidth,page=8]{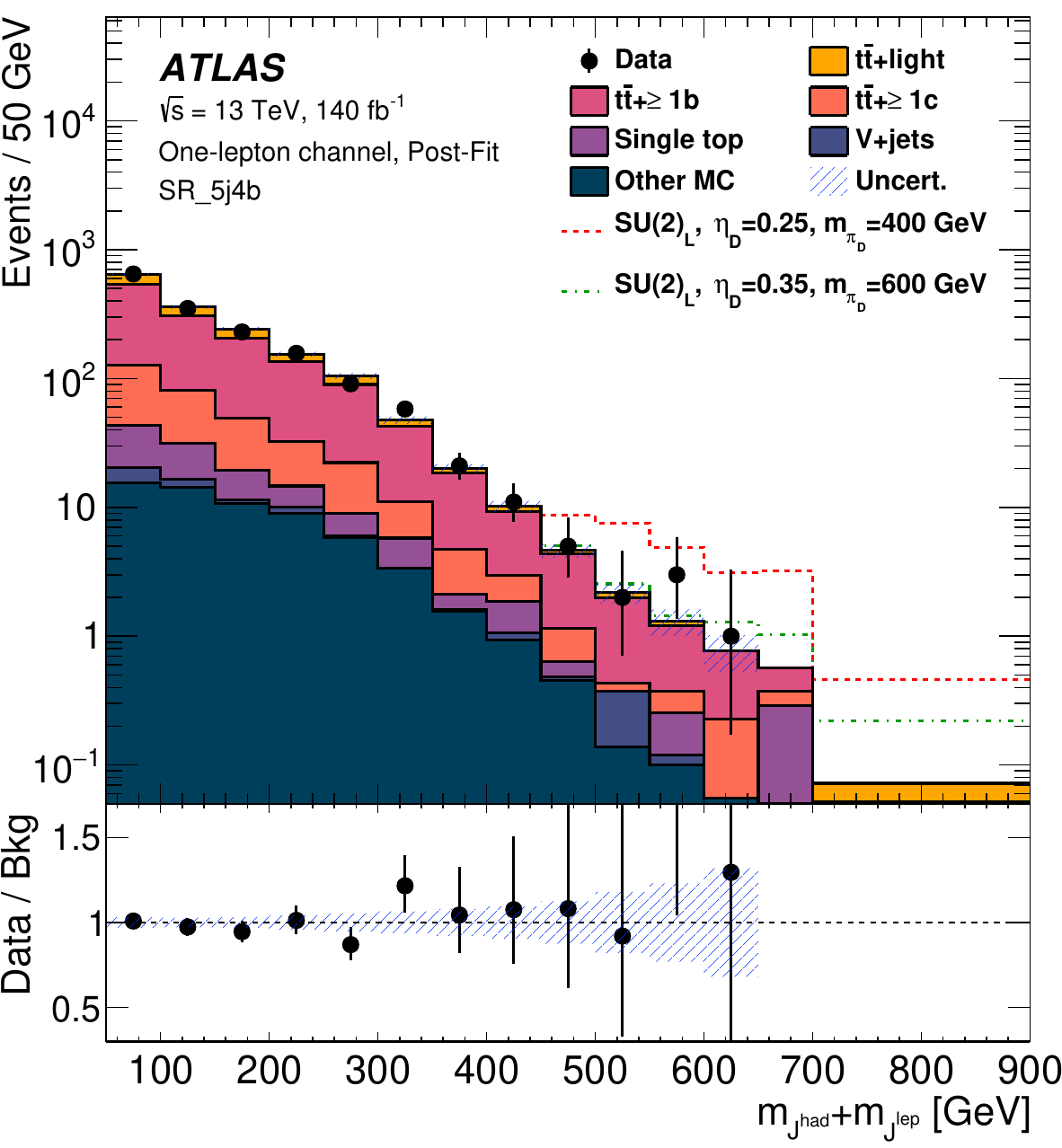}\label{fig:1lep_postfit_SR5j4b}} \\
\subfloat[]{\includegraphics[width=0.35\textwidth,page=9]{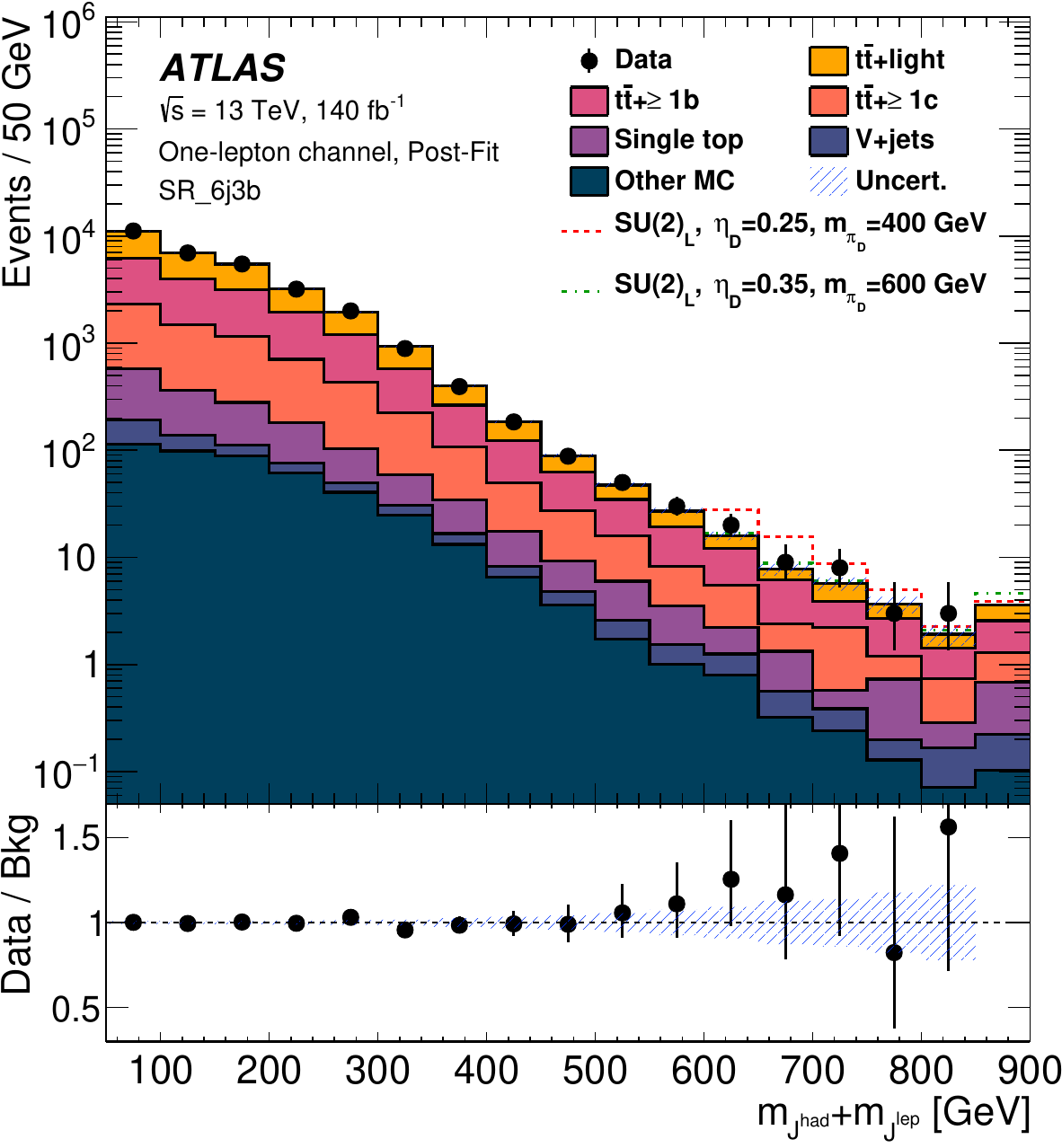}\label{fig:1lep_postfit_SR6j3b}}
\subfloat[]{\includegraphics[width=0.35\textwidth,page=10]{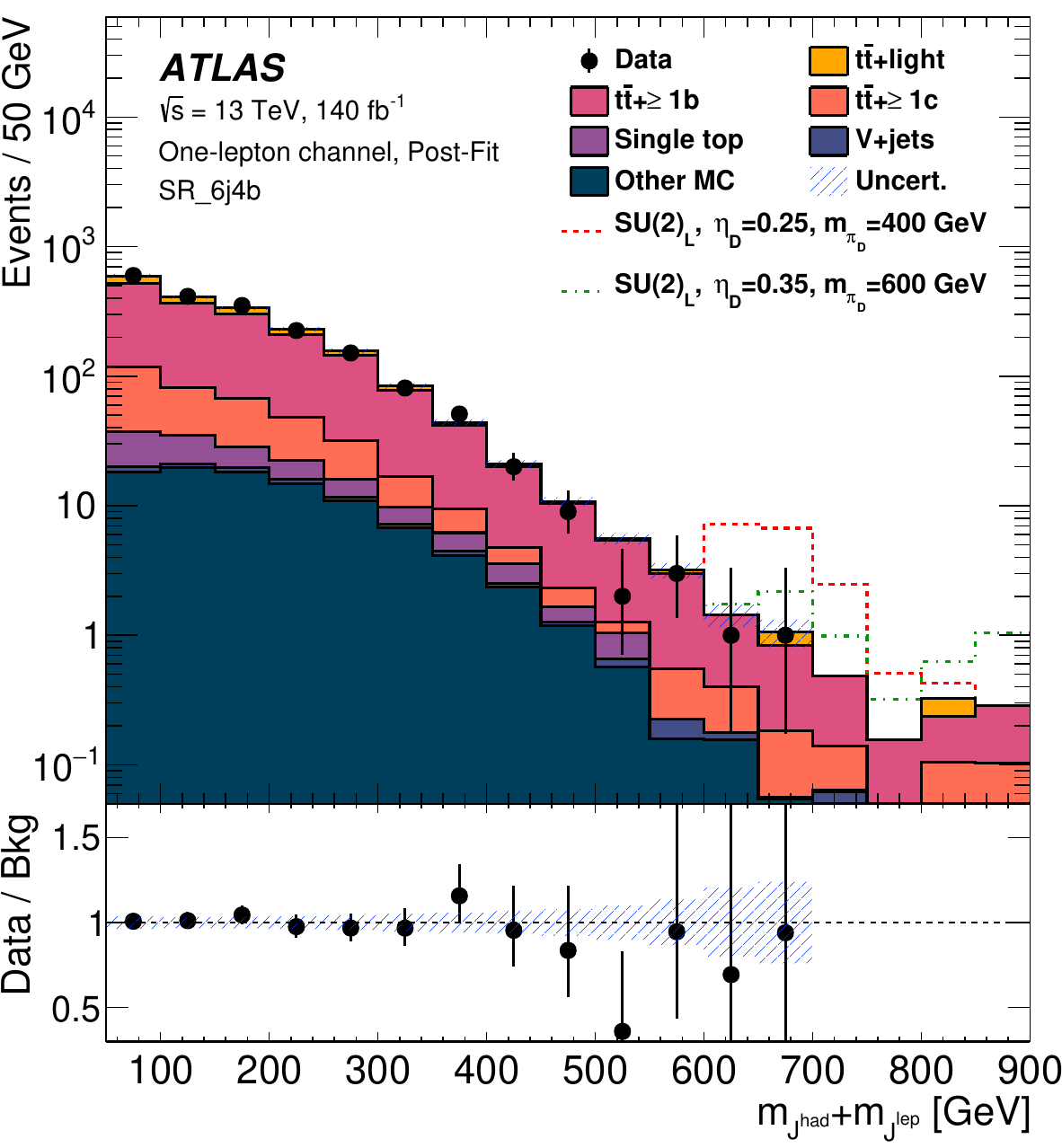}\label{fig:1lep_postfit_SR6j4b}} \\
\subfloat[]{\includegraphics[width=0.35\textwidth,page=11]{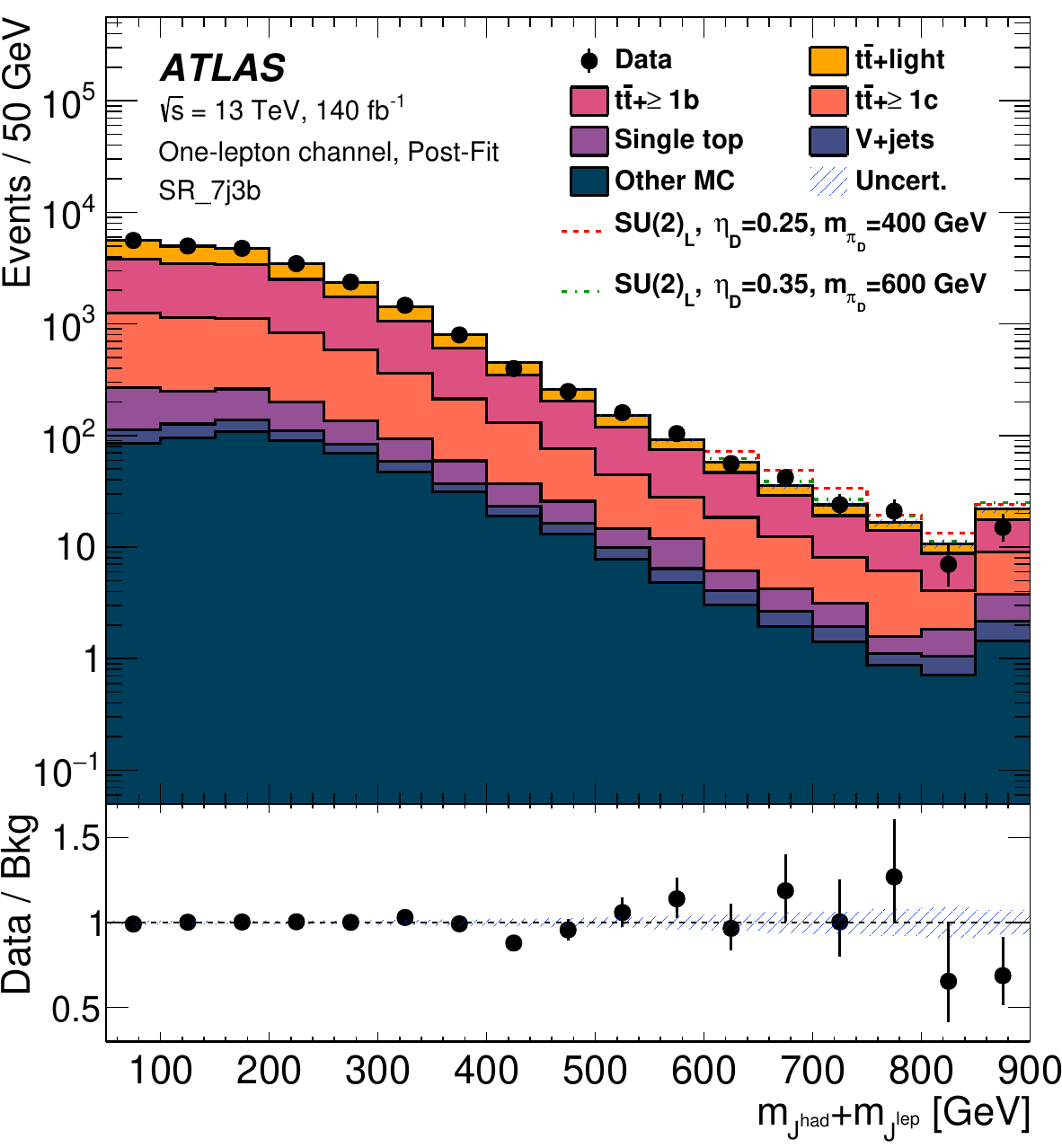}\label{fig:1lep_postfit_SR7j3b}}
\subfloat[]{\includegraphics[width=0.35\textwidth,page=12]{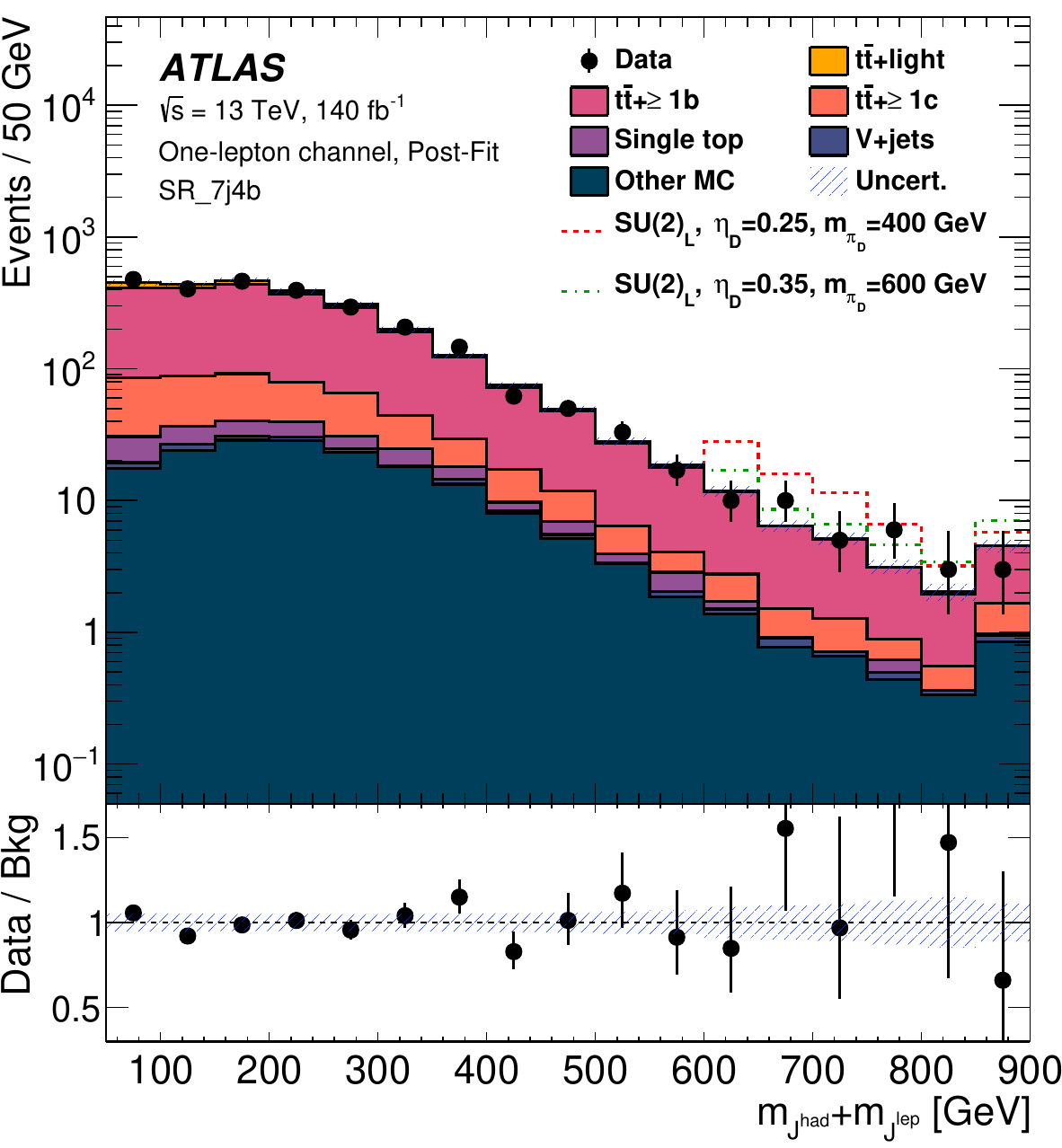}\label{fig:1lep_postfit_SR7j4b}}
\caption{Distributions of \dave in \protect\subref{fig:1lep_postfit_SR5j3b} SR\_5j3b, \protect\subref{fig:1lep_postfit_SR5j4b} SR\_5j4b, \protect\subref{fig:1lep_postfit_SR6j3b} SR\_6j3b, \protect\subref{fig:1lep_postfit_SR6j4b} SR\_6j4b, \protect\subref{fig:1lep_postfit_SR7j3b} SR\_7j3b, and \protect\subref{fig:1lep_postfit_SR7j4b} SR\_7j4b, after the fit to data in all CRs and SRs under the background-only hypothesis.
Two example \sul signal point with $\etaD=0.25$ and $\dpmass=400~\GeV$ is stacked on top of the background. The uncertainty bands contain all statistical and systematic uncertainties. The dashed line in the ratio panel shows the ratio of signal plus background event yields over just background event yields.}
\label{fig:1lep_postfit_SR}
\end{figure}

\begin{table}[htb]
\caption{Observed and predicted event yields after the fit under the background-only hypothesis in the six SR bins in the \onelep channel. The name of each column corresponds to the bins described in Section~\ref{sec:evtsel_onelep}, identified with the number of jets followed by the number of \bjets. The quoted uncertainties contain statistical and systematic components. The total post-fit uncertainty can be smaller than the sum in quadrature of the different components due to correlations resulting from the fit to data.}
\label{tab:1lep_sr_yields}
\centering
\begin{tabular}{c|r@{}lr@{}lr@{}l}
& \multicolumn{2}{c}{SR\_5j3b} & \multicolumn{2}{c}{SR\_5j4b} & \multicolumn{2}{c}{SR\_6j3b} \\
\toprule
\ttb   & 12400$\,\pm\,$&2600 & 1150$\,\pm\,$&180 & 9400$\,\pm\,$&1800 \\
\ttc   & 5900$\,\pm\,$&500 & 230$\,\pm\,$& 40 & 5400$\,\pm\,$&600  \\
\ttlight   & 23100$\,\pm\,$&1900 & 270$\,\pm\,$ &70 & 14300$\,\pm\,$&1200   \\
Single top   & 1800$\,\pm\,$&500 & 59$\,\pm\,$&18 & 1080$\,\pm\,$&330  \\
V+jets   & 274$\,\pm\,$&30 & 10.2$\,\pm\,$&1.5 & 192$\,\pm\,$&22  \\
Other MC   & 416$\,\pm\,$&15 & 61$\,\pm\,$&4 & 465$\,\pm\,$&17 \\
\midrule
Total SM & 10500$\,\pm\,$& 100 & 402$\,\pm\,$&  20 & 8430$\,\pm\,$& 110 \\
\midrule
Data & \multicolumn{2}{c}{10462} & \multicolumn{2}{c}{403} & \multicolumn{2}{c}{8438} \\
\bottomrule
\multicolumn{4}{c}{\phantom{0}}\\
 
& \multicolumn{2}{c}{SR\_6j4b} & \multicolumn{2}{c}{SR\_7j3b} & \multicolumn{2}{c}{SR\_7j4b} \\
\toprule
\ttb & 1210$\,\pm\,$&120 & 9900$\,\pm\,$&2500 & 1900$\,\pm\,$&400 \\
\ttc   & 247$\,\pm\,$&45 & 5400$\,\pm\,$&800 & 350$\,\pm\,$& 90 \\
\ttlight & 220$\,\pm\,$&50 & 8500$\,\pm\,$& 1500 & 180$\,\pm\,$&50  \\
Single top   &  59$\,\pm\,$&18 & 710$\,\pm\,$&220 & 66$\,\pm\,$& 21 \\
V+jets   &  7.7$\,\pm\,$&1.2 & 170$\,\pm\,$&20 & 14.9$\,\pm\,$& 1.9  \\
Other MC   & 99$\,\pm\,$&6 & 620$\,\pm\,$&40 & 187$\,\pm\,$&14 \\
\midrule
Total SM & 477$\,\pm\,$& 22 & 6600$\,\pm\,$& 120 & 640$\,\pm\,$& 40 \\
\midrule
Data & \multicolumn{2}{c}{465} & \multicolumn{2}{c}{6598} & \multicolumn{2}{c}{641} \\
\bottomrule
\multicolumn{4}{c}{\phantom{0}}\\
\end{tabular}
\end{table}


\section{Interpretation}
\label{sec:interpretation}

The results are used to set $95\%$ confidence level (CL) upper limits on the dark pion pair production cross-sections following the $\text{CL}_S$ technique~\cite{Read:2002hq}.
Upper limits on the dark pion production cross-sections are shown in Figure~\ref{fig:allhad_limits_unblinded_su2l_1d} for four slices of \etaD.
The impacts of the different systematic uncertainties, described in Section~\ref{sec:systematics}, on the signal strength, defined as the signal cross-section normalized to the theoretical prediction, of two benchmark signals are summarized in Table~\ref{tab:syst_table}: the analysis is limited by the systematic uncertainty, with dominant contribution from the theoretical sources of uncertainty on the background modeling.
 
\begin{table}[htbp]
\caption{Impact of different categories of systematic uncertainty in the \onelep channel, for two signal benchmarks, relative to the total uncertainty on the fitted signal strength. For each category, the fit is repeated with the corresponding group of nuisance parameters fixed to their best-fit values and the impact for each category is evaluated as the quadrature difference between the signal strength uncertainty in the new fit and in the nominal one, divided by the uncertainty in the nominal fit. The contribution from the statistical uncertainty and the systematic one, further separated into the global instrumental and theoretical uncertainties, are shown. The total systematic uncertainty is different from the sum in quadrature of the different groups due to the correlations among the nuisance parameters in the fit.
}
\label{tab:syst_table}
\centering
\begin{tabular}{l|cc}
& \sul, $\etaD=0.25$, & \sul, $\etaD=0.35$, \\
Category & $m_{\pi_{D}}=400~\GeV$ & $m_{\pi_{D}}=700~\GeV$ \\
\toprule
Luminosity  &  0.03  &0.05\\
Pileup  &  0.05 &0.09 \\
Flavor tagging & 0.28 & 0.26 \\
Leptons & 0.01 & 0.04 \\
Jets  &  0.08  &0.14 \\
\ttb &   0.26  &0.53\\
\ttc  &  0.12  & 0.18\\
\ttlight  &  0.13 & 0.17 \\
Top \pt NNLO reweighting  &  0.08  &0.09 \\
Single top  &  0.06  &0.06\\
\hline
Statistical &   0.28 & 0.24 \\
Instrumental & 0.30& 0.30\\
Theory & 0.38 & 0.63\\
\bottomrule
\end{tabular}
\end{table}
 
Using the predicted dark pion pair production cross-sections, the limits can be translated into limits on dark pion masses in the two-dimensional $\etaD\text{--}\dpmass$ plane. The exclusion contour obtained from the \allhad channel for the \sul model is shown in Figure~\ref{fig:allhad_limits}.
No dark pion masses can be excluded for $\etaD=0.15$.
For $\etaD=0.25$ the exclusion covers the mass range $280~\GeV<\dpmass<520~\GeV$ (expected $280~\GeV<\dpmass<540~\GeV$), while for $\etaD=0.35$, dark pions with masses $\dpmass<430~\GeV$ are excluded (expected $\dpmass<450~\GeV$). 
The exclusion contour obtained from the \onelep channel for \sul is shown in Figure~\ref{fig:1lep_limits} and is observed to fully cover the \allhad limit and significantly extend the probed phase space for this dark meson model. Since the \allhad limit is completely contained within the \onelep limit, there is no expected gain from a combination of the two channels.
Dark pion masses with $\dpmass<940~\GeV$ can be excluded for $\etaD=0.45$, $\dpmass<720~\GeV$ are excluded for $\etaD=0.35$ and for $\etaD=0.25$ the mass region $\dpmass<740~\GeV$ is excluded. The results significantly extend the phase space previously excluded through re-interpretation of other collider searches~\cite{Butterworth:2021jto}.
In the \sur model, cross sections are significantly smaller than for \sul and therefore none of the channels have sensitivity to this model with the current data sample.
 
\begin{figure}[btp]
\centering
\subfloat[]{
\centering
\includegraphics[width=0.445\textwidth]{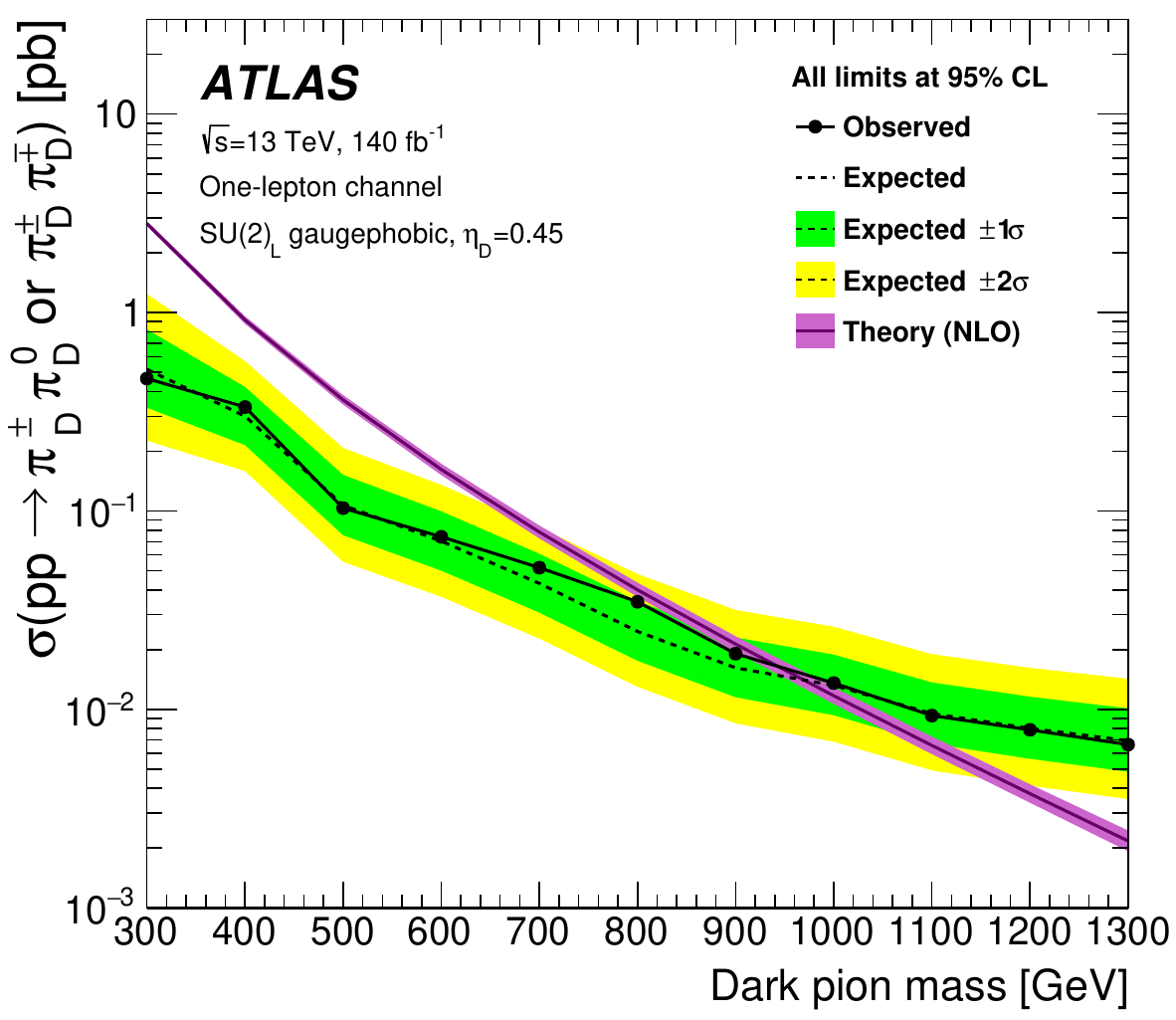}\label{figaux:eta45}
}
\subfloat[]{
\centering
\includegraphics[width=0.445\textwidth]{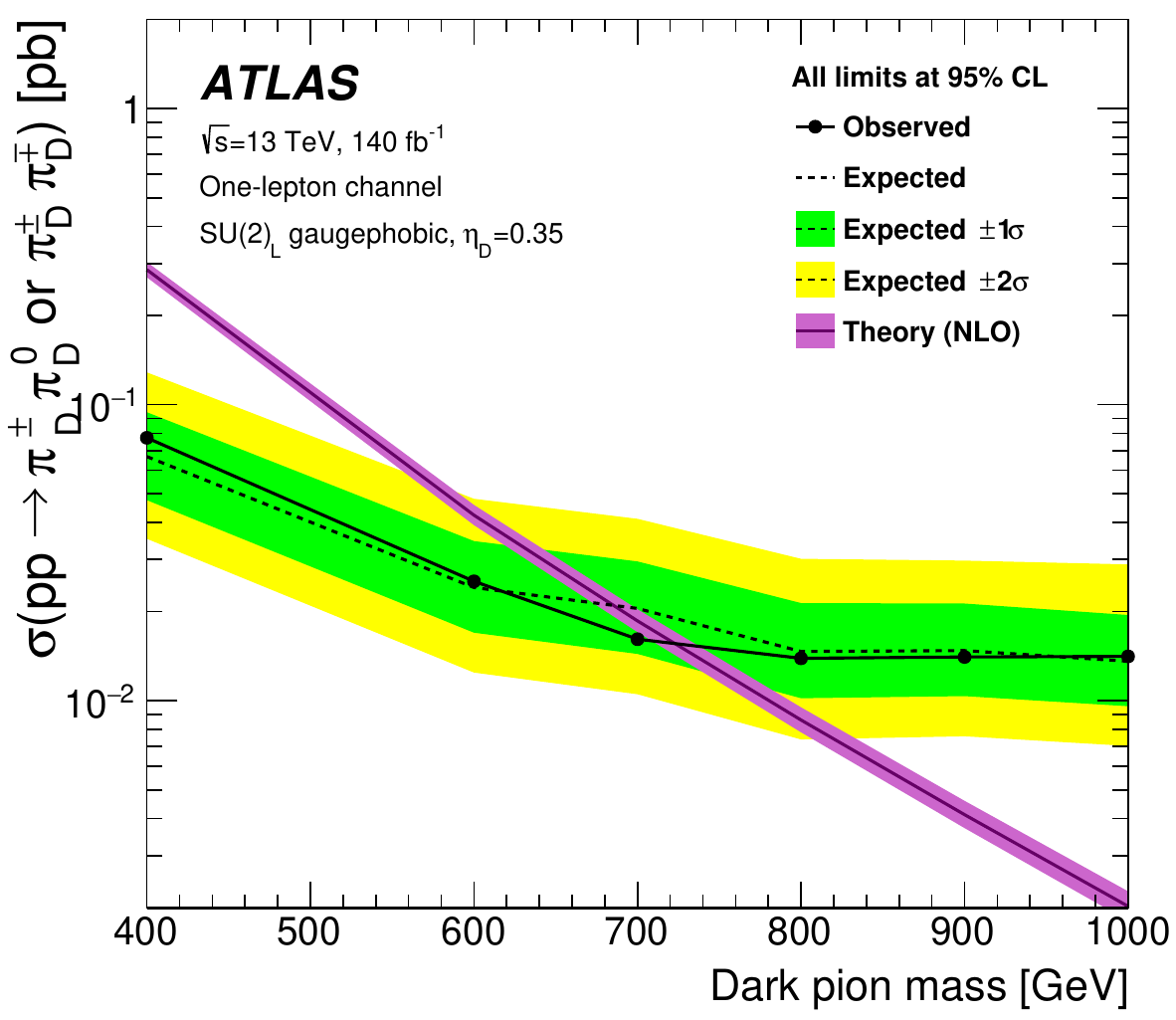}\label{figaux:eta35}
} \\
\subfloat[]{
\centering
\includegraphics[width=0.445\textwidth]{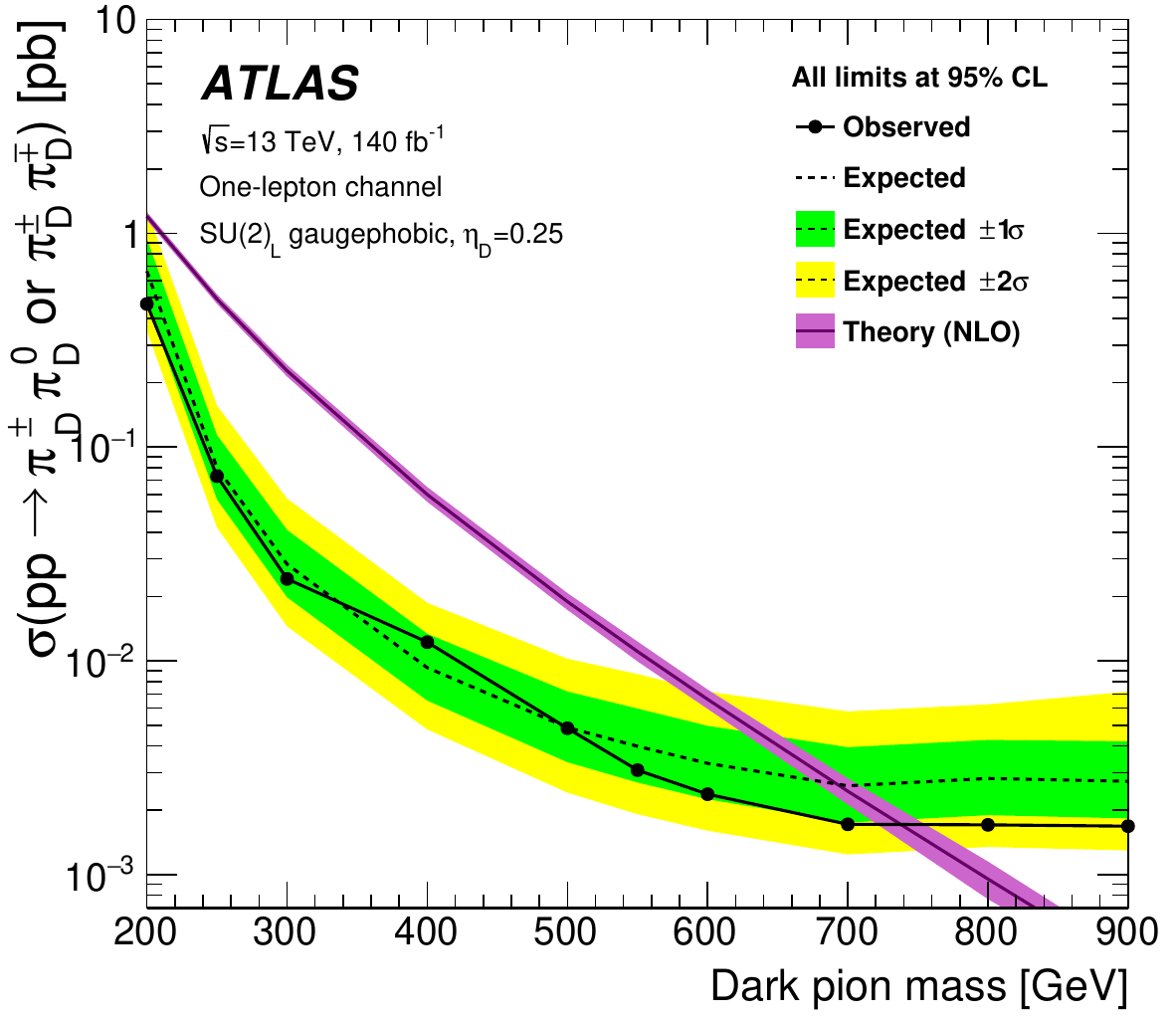}\label{figaux:eta25}
}
\subfloat[]{
\centering
\includegraphics[width=0.445\textwidth]{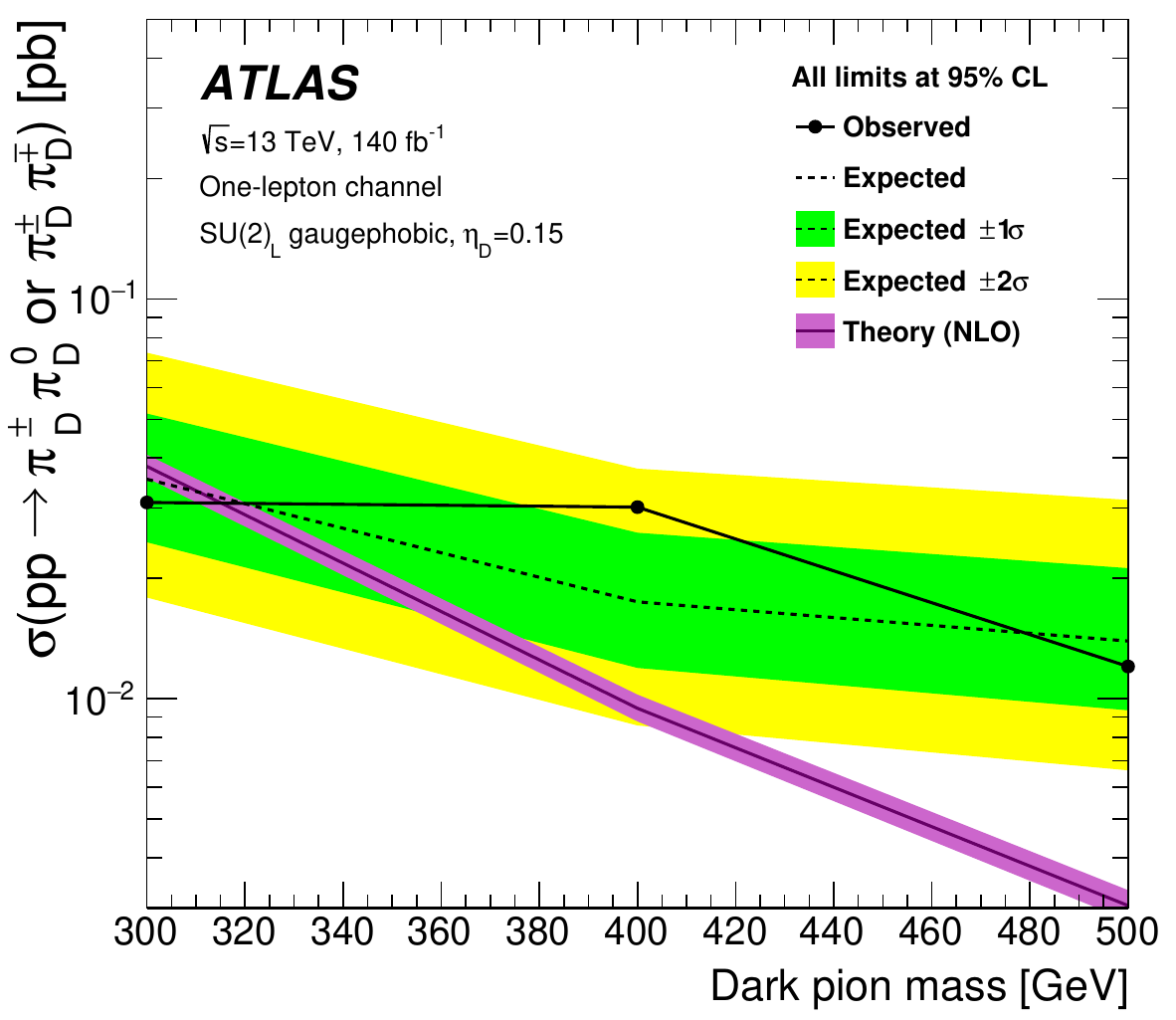}\label{figaux:eta15}
}
\caption{Observed (solid line) and expected (dashed line) limits on the dark pion production cross-section as a function of dark pion mass using the $\text{CL}_S$ method for all \sul models in four slices of $\etaD$: \protect\subref{figaux:eta45} $\etaD$=0.45, \protect\subref{figaux:eta35} $\etaD$=0.35, \protect\subref{figaux:eta25} $\etaD$=0.25, and \protect\subref{figaux:eta15} $\etaD$=0.15. The surrounding shaded bands correspond to one and two standard deviations around the expected limit. The overlaid theory line shows the theoretical dark pion cross-section prediction~\cite{Kribs:2018ilo}.}
\label{fig:allhad_limits_unblinded_su2l_1d}
\end{figure}
 
\begin{figure}[btp]
\centering
\subfloat[]{\includegraphics[width=0.495\textwidth]{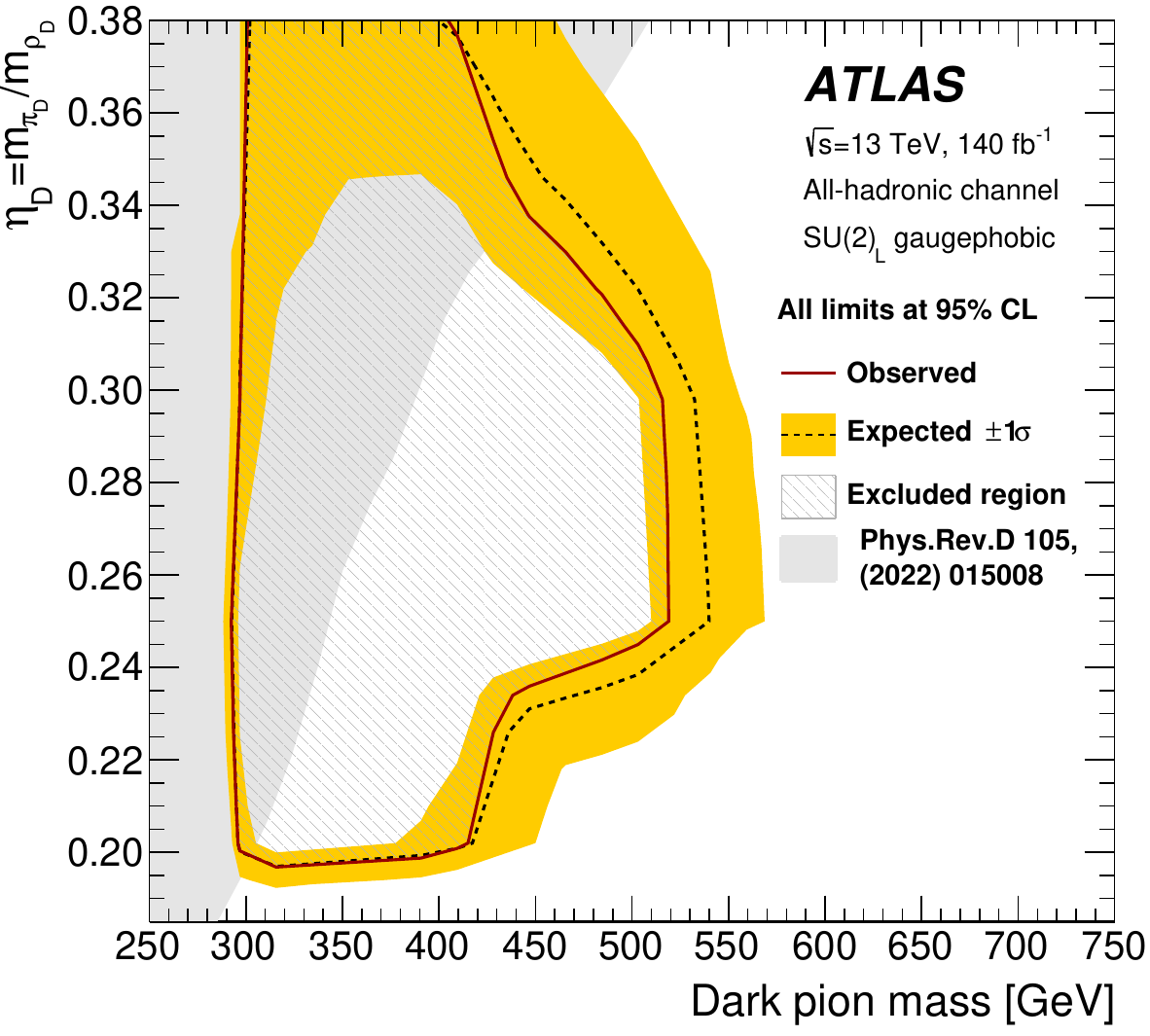}\label{fig:allhad_limits}}
\subfloat[]{\includegraphics[width=0.495\textwidth]{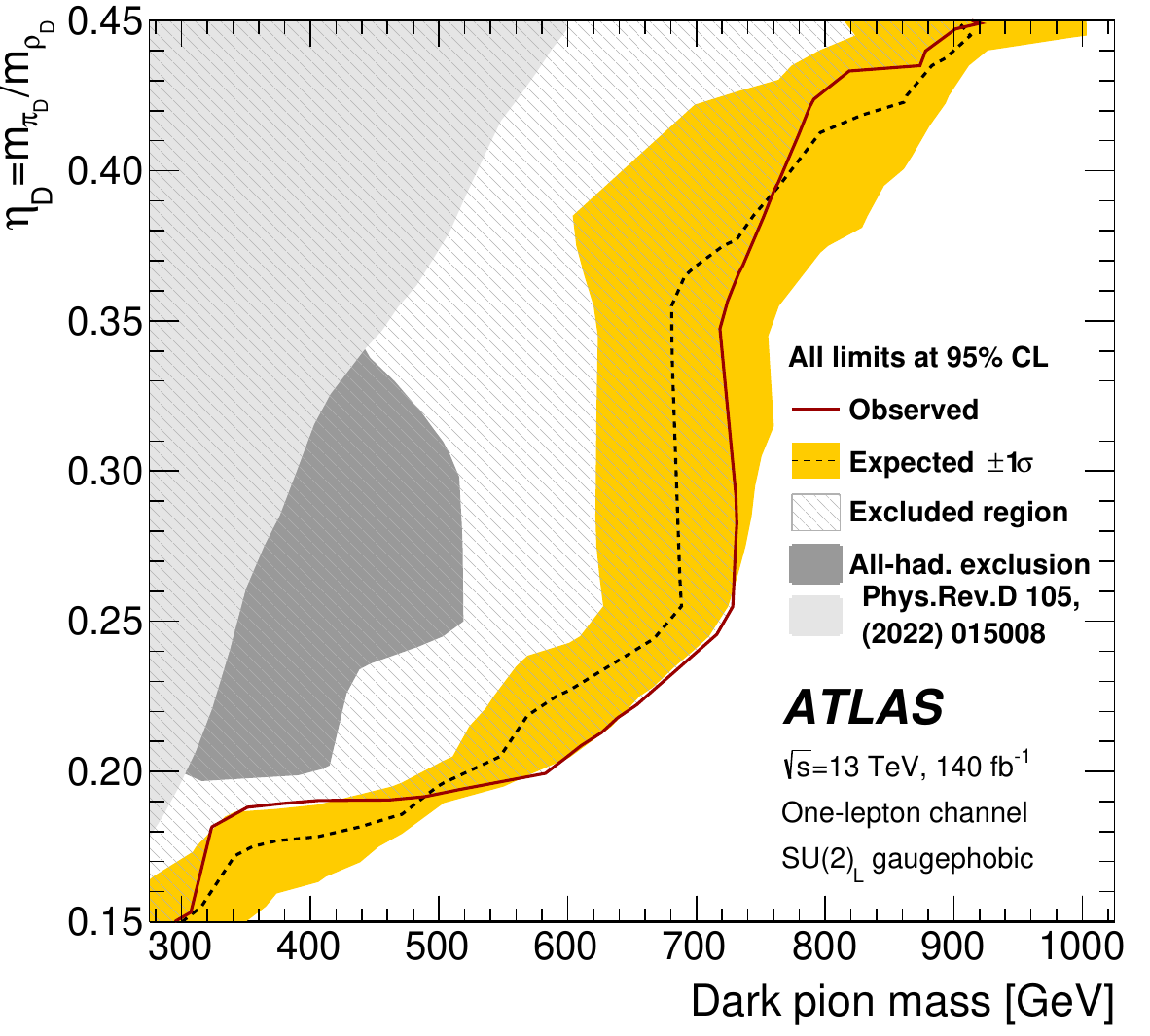}\label{fig:1lep_limits}}
\caption{Observed (solid line) and expected (dashed line) exclusion contours at $95\%$ CL in the $\etaD\text{--}m_{\pi_D}$ plane for \sul signal models in the \protect\subref{fig:allhad_limits} \allhad and \protect\subref{fig:1lep_limits} \onelep channel. Masses that are within the contours are excluded, as indicated by the hatched area. An uncertainty band corresponding to the $\pm 1\sigma$ variation on the expected limit is also indicated. The shaded area in \protect\subref{fig:allhad_limits} and the innermost shaded area in \protect\subref{fig:1lep_limits} indicates the phase space previously excluded through re-interpretation of other collider searches presented in Ref.~\cite{Butterworth:2021jto}. The outermost shaded area in \protect\subref{fig:1lep_limits} indicates the phase space excluded by the analysis in the \allhad channel and is identical to the observed limit shown in \protect\subref{fig:allhad_limits}.}
\label{fig:
limits}
\end{figure}
\FloatBarrier


\FloatBarrier
 
\section{Conclusion}
\label{sec:conclusion}

Results from a search for dark mesons originating from strongly-coupled, $SU(2)$ dark flavor symmetry conserving models and decaying to top and bottom quarks are reported. The analysis is based on the full Run~2 data sample of proton-proton collisions at $\sqrt{s}=13~\TeV$ recorded by the ATLAS detector at the LHC, corresponding to an integrated luminosity of \lumi. The search is performed in the \allhad channel, where the event signature results in eight to ten jets of which at least four originate from bottom quarks, and in the \onelep channel, corresponding to final states with one electron or muon in addition to jets. In the \allhad channel, a data-driven technique is used to estimate the predominant multijet background. In the \onelep channel, the dominant background comes from \ttbar production and is estimated from MC simulations.
 
No excess above the SM background expectation is observed. The strongest exclusion limits are obtained from the \onelep channel. For \sul signals with $m_{\piD}/m_{\rho_D}=0.45$, dark pions with masses $\dpmass<940~\GeV$ are excluded, while for $m_{\piD}/m_{\rho_D}=0.25$ the exclusion covers the mass range $\dpmass<740~\GeV$. These results constitute the first direct collider constraints on this type of model, and significantly extend the phase space previously excluded through re-interpretation of other collider searches.


\section*{Acknowledgments}


We thank CERN for the very successful operation of the LHC and its injectors, as well as the support staff at
CERN and at our institutions worldwide without whom ATLAS could not be operated efficiently.
 
The crucial computing support from all WLCG partners is acknowledged gratefully, in particular from CERN, the ATLAS Tier-1 facilities at TRIUMF/SFU (Canada), NDGF (Denmark, Norway, Sweden), CC-IN2P3 (France), KIT/GridKA (Germany), INFN-CNAF (Italy), NL-T1 (Netherlands), PIC (Spain), RAL (UK) and BNL (USA), the Tier-2 facilities worldwide and large non-WLCG resource providers. Major contributors of computing resources are listed in Ref.~\cite{ATL-SOFT-PUB-2023-001}.
 
We gratefully acknowledge the support of ANPCyT, Argentina; YerPhI, Armenia; ARC, Australia; BMWFW and FWF, Austria; ANAS, Azerbaijan; CNPq and FAPESP, Brazil; NSERC, NRC and CFI, Canada; CERN; ANID, Chile; CAS, MOST and NSFC, China; Minciencias, Colombia; MEYS CR, Czech Republic; DNRF and DNSRC, Denmark; IN2P3-CNRS and CEA-DRF/IRFU, France; SRNSFG, Georgia; BMBF, HGF and MPG, Germany; GSRI, Greece; RGC and Hong Kong SAR, China; ISF and Benoziyo Center, Israel; INFN, Italy; MEXT and JSPS, Japan; CNRST, Morocco; NWO, Netherlands; RCN, Norway; MNiSW, Poland; FCT, Portugal; MNE/IFA, Romania; MESTD, Serbia; MSSR, Slovakia; ARIS and MVZI, Slovenia; DSI/NRF, South Africa; MICIU/AEI, Spain; SRC and Wallenberg Foundation, Sweden; SERI, SNSF and Cantons of Bern and Geneva, Switzerland; NSTC, Taipei; TENMAK, T\"urkiye; STFC/UKRI, United Kingdom; DOE and NSF, United States of America.
 
Individual groups and members have received support from BCKDF, CANARIE, CRC and DRAC, Canada; CERN-CZ, FORTE and PRIMUS, Czech Republic; COST, ERC, ERDF, Horizon 2020, ICSC-NextGenerationEU and Marie Sk{\l}odowska-Curie Actions, European Union; Investissements d'Avenir Labex, Investissements d'Avenir Idex and ANR, France; DFG and AvH Foundation, Germany; Herakleitos, Thales and Aristeia programmes co-financed by EU-ESF and the Greek NSRF, Greece; BSF-NSF and MINERVA, Israel; NCN and NAWA, Poland; La Caixa Banking Foundation, CERCA Programme Generalitat de Catalunya and PROMETEO and GenT Programmes Generalitat Valenciana, Spain; G\"{o}ran Gustafssons Stiftelse, Sweden; The Royal Society and Leverhulme Trust, United Kingdom.
 
In addition, individual members wish to acknowledge support from Armenia: Yerevan Physics Institute (FAPERJ); CERN: European Organization for Nuclear Research (CERN PJAS); Chile: Agencia Nacional de Investigaci\'on y Desarrollo (FONDECYT 1230812, FONDECYT 1230987, FONDECYT 1240864); China: Chinese Ministry of Science and Technology (MOST-2023YFA1605700), National Natural Science Foundation of China (NSFC - 12175119, NSFC 12275265, NSFC-12075060); Czech Republic: Czech Science Foundation (GACR - 24-11373S), Ministry of Education Youth and Sports (FORTE CZ.02.01.01/00/22\_008/0004632), PRIMUS Research Programme (PRIMUS/21/SCI/017); EU: H2020 European Research Council (ERC - 101002463); European Union: European Research Council (ERC - 948254, ERC 101089007), Horizon 2020 Framework Programme (MUCCA - CHIST-ERA-19-XAI-00), European Union, Future Artificial Intelligence Research (FAIR-NextGenerationEU PE00000013), Italian Center for High Performance Computing, Big Data and Quantum Computing (ICSC, NextGenerationEU); France: Agence Nationale de la Recherche (ANR-20-CE31-0013, ANR-21-CE31-0013, ANR-21-CE31-0022, ANR-22-EDIR-0002), Investissements d'Avenir Labex (ANR-11-LABX-0012); Germany: Baden-Württemberg Stiftung (BW Stiftung-Postdoc Eliteprogramme), Deutsche Forschungsgemeinschaft (DFG - 469666862, DFG - CR 312/5-2); Italy: Istituto Nazionale di Fisica Nucleare (ICSC, NextGenerationEU), Ministero dell'Università e della Ricerca (PRIN - 20223N7F8K - PNRR M4.C2.1.1); Japan: Japan Society for the Promotion of Science (JSPS KAKENHI JP22H01227, JSPS KAKENHI JP22H04944, JSPS KAKENHI JP22KK0227, JSPS KAKENHI JP23KK0245); Netherlands: Netherlands Organisation for Scientific Research (NWO Veni 2020 - VI.Veni.202.179); Norway: Research Council of Norway (RCN-314472); Poland: Ministry of Science and Higher Education (IDUB AGH, POB8, D4 no 9722), Polish National Agency for Academic Exchange (PPN/PPO/2020/1/00002/U/00001), Polish National Science Centre (NCN 2021/42/E/ST2/00350, NCN OPUS nr 2022/47/B/ST2/03059, NCN UMO-2019/34/E/ST2/00393, UMO-2020/37/B/ST2/01043, UMO-2021/40/C/ST2/00187, UMO-2022/47/O/ST2/00148, UMO-2023/49/B/ST2/04085); Slovenia: Slovenian Research Agency (ARIS grant J1-3010); Spain: Generalitat Valenciana (Artemisa, FEDER, IDIFEDER/2018/048), Ministry of Science and Innovation (MCIN \& NextGenEU PCI2022-135018-2, MICIN \& FEDER PID2021-125273NB, RYC2019-028510-I, RYC2020-030254-I, RYC2021-031273-I, RYC2022-038164-I), PROMETEO and GenT Programmes Generalitat Valenciana (CIDEGENT/2019/027); Sweden: Swedish Research Council (Swedish Research Council 2023-04654, VR 2018-00482, VR 2022-03845, VR 2022-04683, VR 2023-03403, VR grant 2021-03651), Knut and Alice Wallenberg Foundation (KAW 2018.0157, KAW 2018.0458, KAW 2019.0447, KAW 2022.0358); Switzerland: Swiss National Science Foundation (SNSF - PCEFP2\_194658); United Kingdom: Leverhulme Trust (Leverhulme Trust RPG-2020-004), Royal Society (NIF-R1-231091); United States of America: U.S. Department of Energy (ECA DE-AC02-76SF00515), Neubauer Family Foundation.


\printbibliography

\clearpage
 
\begin{flushleft}
\hypersetup{urlcolor=black}
{\Large The ATLAS Collaboration}

\bigskip

\AtlasOrcid[0000-0002-6665-4934]{G.~Aad}$^\textrm{\scriptsize 104}$,
\AtlasOrcid[0000-0001-7616-1554]{E.~Aakvaag}$^\textrm{\scriptsize 17}$,
\AtlasOrcid[0000-0002-5888-2734]{B.~Abbott}$^\textrm{\scriptsize 123}$,
\AtlasOrcid[0000-0002-0287-5869]{S.~Abdelhameed}$^\textrm{\scriptsize 119a}$,
\AtlasOrcid[0000-0002-1002-1652]{K.~Abeling}$^\textrm{\scriptsize 56}$,
\AtlasOrcid[0000-0001-5763-2760]{N.J.~Abicht}$^\textrm{\scriptsize 50}$,
\AtlasOrcid[0000-0002-8496-9294]{S.H.~Abidi}$^\textrm{\scriptsize 30}$,
\AtlasOrcid[0009-0003-6578-220X]{M.~Aboelela}$^\textrm{\scriptsize 45}$,
\AtlasOrcid[0000-0002-9987-2292]{A.~Aboulhorma}$^\textrm{\scriptsize 36e}$,
\AtlasOrcid[0000-0001-5329-6640]{H.~Abramowicz}$^\textrm{\scriptsize 154}$
% ...
\bigskip
\\

% ...
$^{17}$Department for Physics and Technology, University of Bergen, Bergen; Norway.\\
% ...
$^{30}$Physics Department, Brookhaven National Laboratory, Upton NY; United States of America.\\
% ...
$^{36}$$^{(a)}$Facult\'e des Sciences Ain Chock, Universit\'e Hassan II de Casablanca;$^{(b)}$Facult\'{e} des Sciences, Universit\'{e} Ibn-Tofail, K\'{e}nitra;$^{(c)}$Facult\'e des Sciences Semlalia, Universit\'e Cadi Ayyad, LPHEA-Marrakech;$^{(d)}$LPMR, Facult\'e des Sciences, Universit\'e Mohamed Premier, Oujda;$^{(e)}$Facult\'e des sciences, Universit\'e Mohammed V, Rabat;$^{(f)}$Institute of Applied Physics, Mohammed VI Polytechnic University, Ben Guerir; Morocco.\\
% ...
$^{45}$Physics Department, Southern Methodist University, Dallas TX; United States of America.\\
% ...
$^{50}$Fakult\"{a}t Physik , Technische Universit{\"a}t Dortmund, Dortmund; Germany.\\
% ...
$^{56}$II. Physikalisches Institut, Georg-August-Universit\"{a}t G\"ottingen, G\"ottingen; Germany.\\
% ...
$^{104}$CPPM, Aix-Marseille Universit\'e, CNRS/IN2P3, Marseille; France.\\
% ...
$^{119}$$^{(a)}$New York University Abu Dhabi, Abu Dhabi;$^{(b)}$United Arab Emirates University, Al Ain; United Arab Emirates.\\
% ...
$^{123}$Homer L. Dodge Department of Physics and Astronomy, University of Oklahoma, Norman OK; United States of America.\\
% ...
$^{154}$Raymond and Beverly Sackler School of Physics and Astronomy, Tel Aviv University, Tel Aviv; Israel.\\
% ...
$^{a}$ Also Affiliated with an institute covered by a cooperation agreement with CERN.\\
$^{b}$ Also at An-Najah National University, Nablus; Palestine.\\
$^{c}$ Also at Borough of Manhattan Community College, City University of New York, New York NY; United States of America.\\
$^{d}$ Also at Center for Interdisciplinary Research and Innovation (CIRI-AUTH), Thessaloniki; Greece.\\
$^{e}$ Also at CERN, Geneva; Switzerland.\\
$^{f}$ Also at CMD-AC UNEC Research Center, Azerbaijan State University of Economics (UNEC); Azerbaijan.\\
% ...

\end{flushleft}


\end{document}